\documentclass[a4paper,12pt]{article}
\pdfoutput=1
\pdfoutput=1
\usepackage{jheppub,undertilde}
\usepackage{amsthm}
\usepackage{graphicx}
\usepackage{color}
\usepackage{graphicx,textcomp,float,gensymb,wrapfig, enumitem,comment,dsfont,subfigure,framed,slashed,appendix,wrapfig}
 
 \newcommand{\be}{\begin{equation}}
\newcommand{\ee}{\end{equation}}
\def\bsp#1\esp{\begin{split}#1\end{split}}

\newcommand{\bea}{\begin{eqnarray}}  
\newcommand{\eea}{\end{eqnarray}}  
\def\code#1{\texttt{#1}}

\preprint{FTUV-16-0419, IFIC/16-32\\}
\title{Sterile neutrino portal to Dark Matter I: \\ The $U(1)_{B-L}$ case}
\author[a]{Miguel Escudero,}
\author[a]{Nuria Rius}
\author[b]{and Ver\'onica Sanz}
\affiliation[a]{Departamento de F\'isica Te\'orica and IFIC, Universidad de Valencia-CSIC,
C/ Catedr\'atico Jos\'e Beltr\'an, 2, E-46980 Paterna, Spain}
\affiliation[b]{Department of Physics and Astronomy, University of Sussex, Brighton BN1 9QH, UK\\}

\abstract{
In this paper we explore the possibility that the sterile neutrino and Dark Matter sectors in the Universe have a common origin. We study the consequences of this assumption in the simple case of coupling the dark sector to the Standard Model via a global $U(1)_{B-L}$, broken down spontaneously by a dark scalar. This dark scalar provides masses to the dark fermions and communicates with the Higgs via a Higgs portal coupling. We find an interesting interplay between Dark Matter annihilation to dark scalars  -  the CP-even that mixes with the Higgs and the CP-odd which becomes a Goldstone boson, the Majoron - and heavy neutrinos, as well as collider probes via the coupling to the Higgs.  
Moreover, Dark Matter annihilation into sterile neutrinos and its subsequent decay to gauge bosons and quarks, charged leptons or neutrinos lead to indirect detection signatures which are close to current bounds on the 
gamma ray flux from the galactic center and dwarf galaxies.}

\emailAdd{miguel.escudero@ific.uv.es}
\emailAdd{nuria.rius@ific.uv.es}
\emailAdd{v.sanz@sussex.ac.uk}

\keywords{}

\begin{document}
\maketitle
\flushbottom

\section{Introduction}
The study of the dark Universe is one of the best handles to understand what lies beyond the Standard Model (SM), particularly possible connections between Dark Matter and other sectors. The SM neutrino sector is especially interesting, as the observation of neutrino masses already points to new physics beyond the SM,
possibly in the form of massive right-handed neutrinos. This raises the question whether these two new forms of massive particles, Dark Matter and right-handed neutrinos, are somewhat linked.

A very minimal possibility would be that of right-handed neutrinos constituting the Dark Matter of the Universe~\cite{Dodelson:1993je}. Yet, this option is tightly constrained in a region of small mixing with active neutrinos and mass around the keV, which will be explored in upcoming experiments and potentially excluded, see~\cite{Adhikari:2016bei} for a recent review on the subject. 

In this paper we propose a different scenario, where sterile neutrinos and a fermionic Dark Matter particle would have a common origin within a dark sector. These dark fermions would exhibit couplings to a dark scalar, which would bring a source of Majorana masses. The right-handed neutrinos would mix with active neutrinos, providing a link to the SM, which Dark Matter would inherit via exchanges of the dark scalar. Additionally, the dark scalar could couple to the SM via a Higgs portal, providing Dark Matter yet another mechanism to communicate with the SM. In this paper we choose the rather natural option of charging the dark sector under $U(1)_{B-L}$, but another minimal choice would be 
to assume an exact symmetry of the dark sector which stabilizes the lightest dark particle and allows 
 to communicate with the SM via the right-handed neutrinos, singlets under both the SM and the dark 
 group, see~\cite{Macias:2015cna,Gonzalez-Macias:2016vxy} and~\cite{Escudero:2016ksa}. 

The paper is organized as follows. After presenting the set-up of our model in Sec.~\ref{sec:setup}, and the consequences of the breaking of $U(1)_{B-L}$ in the scalar sector in Sec.~\ref{sec:physical}, we move onto the phenomenology of the model in Sec.~\ref{sec:pheno}. 
The study of Higgs decays and direct detection in Secs.~\ref{sec:Hinv} and~\ref{sec:DD}, does lead to strong contraints on the mixing between the dark scalar and the Higgs. 
How Dark Matter can satisfy the observed relic abundance is explored in Sec.~\ref{sec:relic}, and the correlation with indirect detection in Sec.~\ref{sec:ID}.
We discuss the implications of a strongly self-interacting Dark Matter in this model in Sec.~\ref{sec:SIDM}, just before moving onto summarizing our findings in Sec.~\ref{sec:results}. 
We conclude in Sec.~\ref{sec:concls} by providing a summary of the results and outlook of possible new directions of investigation.

\section{A Dark sector with $U(1)_{B-L}$}\label{sec:setup}
We consider the following set-up: we extend the SM with a complex scalar field, $\phi$ and  $n$ chiral (RH) fermion fields, $\Psi_R$.  All these new fields are SM singlets, and charged under a global $U(1)$ symmetry which 
can be identified with $U(1)_{B-L}$, so that $L_\phi =2$ and $L_{\Psi_R} = 1$~\footnote{Note that $U(1)_{B-L}$ is the only anomaly-free global symmetry in the SM. Therefore, extensions of the SM including a {\it gauged} $U(1)_{B-L}$ have been considered in various contexts, and in particular in scenarios where the breaking appears at low-scale (e.g.~\cite{Chikashige:1980ui,Khalil:2006yi,Iso:2009ss,Kanemura:2014rpa}).}. 
\begin{figure}[h!]
\centering
\includegraphics[scale=0.2]{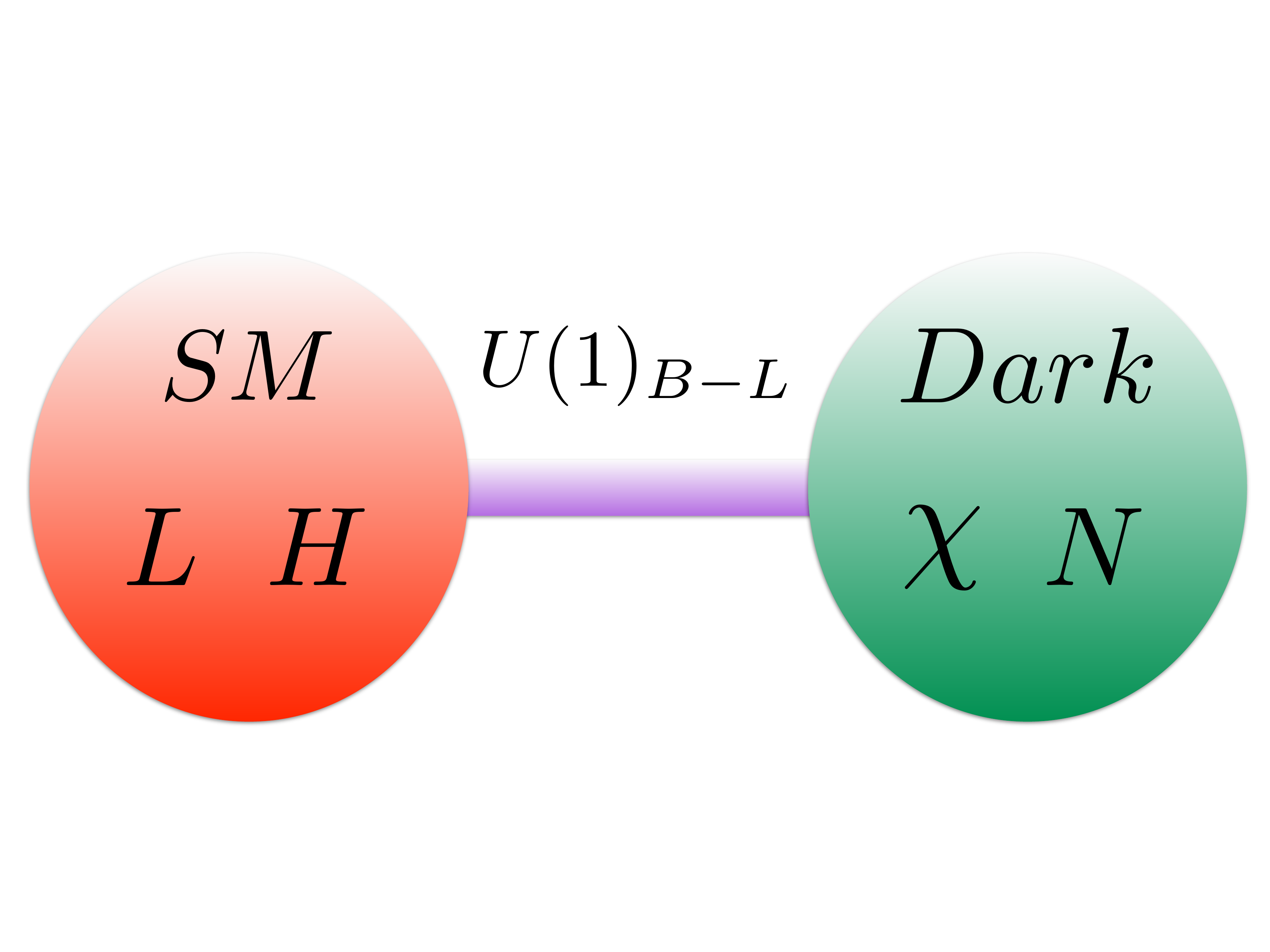}
\end{figure}

Moreover, we assume that (for the reasons explained below) some of the dark fermions  have  vanishing or suppressed  
coupling to the SM singlet operator $\overline {L}_L H$, so they could be stable (or cosmologically stable);
we will denote such stable fermion(s) by $\chi_R$, as opposed to the rest of the dark fermions, which we will call $N_{R}$.

Communication between the Standard Model fields and the new singlet sector $(\phi,\Psi_R)$ is determined by the $U(1)_{B-L}$ charges and the requirement of renormalizability of the interactions. 
The relevant part of the Lagrangian reads: 
\bea
\label{eq:lag1}
{\cal L} &\supset& \mu_H^2 H^\dagger H - \lambda_H (H^\dagger H)^2  + 
\mu_\phi^2 \phi^\dagger \phi - \lambda_\phi (\phi^\dagger \phi)^2 - 
\lambda_{H\phi} (H^\dagger H) \, (\phi^\dagger \phi)  
\\
&-& \left( \frac{\lambda_{\chi a b}}{\sqrt{2}} \phi \, \overline{\chi}_{Ra} \chi_{Rb}^c + h.c. \right) - 
\left( \frac{\lambda_{N a b}}{\sqrt{2}} \phi \, \overline{N}_{Ra} N_{Rb}^c + h.c. \right) 
- (Y_{\alpha a} \overline {L}_L^\alpha H N_{Ra}  + h.c.)
\nonumber 
\eea
where $\alpha=e,\mu, \tau$ denotes lepton flavour, $a,b$ refers to the different dark fermion species 
and the Yukawa coupling matrices $\lambda_\chi, \lambda_N$ are symmetric.

The  coupling between the Higgs and the complex scalar $\phi$, $\lambda_{H\phi}$, is a generalized {\it Higgs portal} coupling, whereas the direct coupling between the right-handed fermions $N_R$ and the SM via the mixing term $Y_{\alpha a}$ leads to masses for the active neutrinos. The Yukawa coupling between the dark scalar and dark fermions $\lambda_\chi, \lambda_N$ generates a Majorana mass for the $\chi$ fields and sterile neutrinos 
 provided $\phi$ develops a vev. Details on neutrino mass generation in this set-up can be found in Sec.~\ref{sec:nmass}.

So far, we have described a new sector linked to the origin of neutrino masses. We now consider whether this sector could also describe Dark Matter. In our set-up there are two possible candidates {\it 1.)} 
the right-handed fermions $\chi_{Ra}$, and/or {\it 2.)} a component of the complex scalar field $\phi$. We discuss in turn each possibility. 

{\bf Fermionic Dark Matter:} 

Possible mechanisms to ensure stability of the dark fermions $\chi_R$ could be:

\begin{enumerate}
 \item {\it $Z_2$ symmetry:} The simplest possibility is that  the fermions 
$\chi_R$ are odd under an exact $Z_2$ symmetry, 
  while all the SM particles, the singlet scalar $\phi$ and the remaining fermions $N_R$ are even.
Then, the   Yukawa coupling of $\chi_R$ to SM leptons will be forbidden, resulting on a stable 
sterile neutrino Dark Matter.  
  
  \item {\it Compositeness:} The dark sector is a low-energy description of a new strongly coupled sector (charged under the global $U(1)_{B-L}$), with the dark particles bound states of the strong dynamics. Mixing between the SM operator $ \overline {L}_L^\alpha H$ and fermionic bound states ${\cal O}^a$ with lepton number are allowed, but the strength of this mixing is determined by the anomalous dimension of ${\cal O}^a$. One could also describe this set-up in terms of a holographic dual, where operators from a strongly coupled sector like ${\cal O}^a$ are represented by states living in more than 4D, ${\cal O}^a(x) \to \chi_R(x,z)$, with $z$ is the extra dimensional coordinate. In this holographic picture, the SM particles (lepton doublets, Higgs) are localized at some distance from where the fields $\chi_R$ have their main support. The values of $Y_{\chi \alpha a}$ are obtained via dimensional reduction from 5D to 4D, namely computing overlaps of the wavefunctions of the Higgs, lepton doublets and dark fermions~\cite{ArkaniHamed:1998vp, ArkaniHamed:1999dc}
 \bea
 Y_{\chi \alpha a} \propto \int d z f_H (z) f_{L^\alpha} (z) f_{\chi_{Ra}} (z) \ .
 \eea
 In warped geometries, ${\cal O}(1)$ differences in localization parameters can lead to exponential hierarchies among the different entries in $Y_{\chi \alpha a}$~\cite{Grossman:1999ra} and hence (meta)stability of some dark fermions.
 
  \item {\it Exotic lepton number:} If there are at least two Weyl fermions in the dark sector, 
  they could have lepton number different from $\pm1$, so that the 
  Yukawa interaction $\overline {L}_L^\alpha H \chi_{Ra,b}$ is forbidden but the coupling 
  $\phi \, \overline{\chi}_{Ra} \chi_{Rb}^c$ is allowed provided $L_a + L_b = - 2$.
  This scenario leads to Dirac Dark Matter particle, and has been explored in 
  \cite{Lindner:2011it} in the context of the Zee-Babu model for neutrino masses.
  
   \item {\it Different dark sector representations:} The dark sector could have further symmetry 
   structure (more complex than the simple $Z_2$ symmetry described above), so that some of the chiral fermions are singlets under the dark symmetry group and thus can couple to $\overline {L}_L^\alpha H$, while $\chi_{Ra}$ may transform non 
   trivially under the dark group. Then,  $\phi \, \overline{\chi}_{Ra} \chi_{Ra}^c$ is
   invariant  and thus allowed, but the Yukawa coupling with the SM fermions is forbidden 
   by the dark symmetry. 
   
   \end{enumerate}
 
 In this paper we assume that either mechanism 1 or 3 are at work, and we discuss the 
 phenomenology of these two minimal realizations. In the case of an additional dark sector symmetry, 
 if it is global the only difference will be an extra factor in the annihilation cross sections of Sec.~\ref{sec:relic}, related to the dimension of the representation to which $\chi_R$ belongs, so our results can be easily 
 re-scaled; however, if the dark symmetry is gauged, Dark Matter self-interactions could modify some of our  findings. 

{\bf Scalar Dark Matter:} 
The imaginary part of the complex field (the so-called Majoron, $\eta$) could be a Dark Matter candidate~\cite{Chikashige:1980ui,Schechter:1981cv} provided it acquires a mass, e.g. via non-perturbative gravitational effects which break the global symmetry~\cite{Coleman:1988tj,Akhmedov:1992hi}. For a recent review on the subject see \cite{Lattanzi:2014mia}. 

A massive Majoron decays at tree level to a pair of light neutrinos with a rate that scales as
\cite{Akhmedov:1992hi}:
\be
\Gamma(\eta  \to \nu \nu) = \frac{m_\eta}{8 \pi} \left(\frac{m_\nu}{v_\phi}\right)^2   \ ,
\ee
where $m_\nu$ is the mass scale of ordinary neutrinos and $v_\phi$ the scale of $U(1)_{B-L}$ 
spontaneous breaking. Therefore, for instance if $m_\eta \lesssim 10$ keV and 
 $v_\phi \gtrsim 10^8$ GeV, the lifetime of the Majoron can be large enough 
for it to be stable on cosmological scales, while for $v_\phi$ in the TeV range the Majoron decays very fast.  

Moreover, the massive Majoron might also decay into two photons at the loop level, 
although this mode is model-dependent. 
While it does not occur in the minimal singlet Majoron scenario that we are considering, it is induced at one 
loop in the more general seesaw model which includes also a triplet scalar field coupled to the SM lepton 
doublets \cite{Bazzocchi:2008fh}, and it could also be present if the dark sector contains other chiral fermion representations charged under the SM gauge group with masses of order $\Lambda  \gg v_\phi$, which 
would make the global symmetry $U(1)_{B-L}$ anomalous. 
Current experimental bounds on pseudo-Goldstone bosons with ${\rm BR}(\eta \to \gamma \gamma) \sim 1$ 
imply that it can have a lifetime longer than $\sim 10^{20}$ years if its mass is $m_\eta \lesssim 100$ keV, 
while for heavier masses, $m_\eta \gtrsim 10$ MeV, the lifetime has to be shorter than one minute 
 \cite{Garcia-Cely:2013wda}.

Both Dark Matter candidates, a keV scale sterile neutrino and a massive Majoron 
have received much attention in the literature, so 
 we do not explore such possibilities any further in this paper. Instead, we focus on the fermionic Dark Matter 
 scenario extending the study to larger masses, in the typical WIMP range, which to our 
 knowledge has not been considered up to now. 
 It has been studied in the framework of 
 gauged $U(1)_{B-L}$, however then there is also a new $Z'$ gauge boson and constraints from direct 
 searches set a lower bound on the scale of $U(1)_{B-L}$ symmetry breaking of order  few TeV 
 \cite{Carena:2004xs}. 
 As a consequence, the correct Dark Matter relic abundance can only be obtained near the resonance regions, when twice the Dark Matter mass is approximately equal to the mass of any of the mediators \cite{Okada:2010wd}.
 
 Notice that our scenario differs from one without spontaneous symmetry breaking in several ways.
 In the absence of the  $U(1)_{B-L}$ global symmetry, masses for the 
 Dark Matter and sterile neutrinos, which explicitly break lepton number, should be added by hand instead of being a consequence of the breaking of $U(1)_{B-L}$ . Therefore in that case
 they would be independent of their corresponding couplings to the dark scalar $\phi$, 
 while in our case they are related by the vev of $\phi$ (see Sec.~\ref{sec:physical}).
 Moreover, there would not be  a Goldstone boson, and one would expect  the real and the  imaginary components of  the scalar 
 $\phi$ to have masses of the same order.

 \section{Parametrization of the physical states}\label{sec:physical}
Both the SM Higgs and the complex scalar $\phi$ can develop vevs, which would break the symmetry 
group $SU(2)_L \times U(1)_Y \times U(1)_{B-L} \rightarrow U(1)_{em} \times Z_2$. 
We parametrize the scalar sector as:

\be
H = \left( \begin{array}{c}
G^+ \\
\frac{v_H + \tilde h + iG^0}{\sqrt{2}}
\end{array}
\right)  \ , \qquad \phi = \frac{v_\phi + \tilde \rho + i \eta}{\sqrt{2}} \ ,
\ee
where $v_H = 245$ GeV. 
The minimization of the scalar potential in eq.~(\ref{eq:lag1}), leads to the following 
tree-level relations between the Lagrangian parameters and the vacuum expectation values 
of the fields $H,\phi$:
\be
\mu_H^2 = \lambda_H  v_H^2 + \frac 1 2 \lambda_{H \phi} v_\phi^2 \ , 
\qquad
\mu_\phi^2 = \lambda_\phi  v_\phi^2 + \frac 1 2 \lambda_{H \phi} v_H^2 \ .
\ee
The scalar sector then contains two CP even massive real scalars, $\tilde h$ and $\tilde \rho$ 
which mix and upon diagonalization of the mass matrix lead to the mass eigenstates 
$h$ and $\rho$, with masses $m_h$ and $m_\rho$, respectively. The state $h$ is identified
with the scalar boson of $m_h =125$ Gev discovered at the LHC.  

The masses of the physical states are:
\bea
\label{eq:ms}
m_h^2 &=& 2 \lambda_H v_H^2 \cos^2 \theta + 2 \lambda_\phi v_\phi^2 \sin^2\theta 
-  \lambda_{H\phi}    v_H v_\phi \sin 2 \theta
\\
m_\rho^2 &=& 2 \lambda_H v_H^2 \sin^2 \theta + 2 \lambda_\phi v_\phi^2 \cos^2\theta 
+  \lambda_{H\phi}    v_H v_\phi \sin 2 \theta
\eea
and the mixing angle 
\be
\label{eq:mixa}
\tan 2\theta = \frac{\lambda_{H\phi}  v_H v_\phi }{\lambda_\phi v_\phi^2 - \lambda_H v_H^2}
\ee
There is also a CP odd massless real
scalar $\eta$, which is the Goldstone boson of the spontaneous breaking of the global $U(1)_{B-L}$ 
symmetry, the Majoron \cite{Chikashige:1980ui}.
We assume that, even if quantum gravity effects break the global $U(1)_{B-L}$  and provide a mass to
the Majoron, it is much lighter than the other dark particles, i.e., $m_\eta \ll {\cal O}({\rm GeV})$ and we  neglect it in our analysis.

The quartic couplings in the Lagrangian can be written in terms of the physical masses and the mixing angle in the CP even scalar sector as follows:
\bea
\lambda_H &=& \frac{m_h^2 \cos^2 \theta + m_\rho^2 2 v_H^2}{2 v_H^2}
\nonumber \\
\lambda_\phi &=& \frac{m_h^2 \sin^2 \theta + m_\rho^2 \cos^2\theta }{2 v_\phi^2}
\\
\lambda_{H\phi} &=& \frac{( m_\rho^2 - m_h^2) \sin 2 \theta }{2 v_H v_\phi}
\nonumber
\eea

Regarding the (neutral) lepton sector of the model, let us 
denote $\chi$ the fermion without Yukawa coupling to the SM lepton doublets, i.e., the Dark Matter candidate, 
and $N_a$ the fermions with couplings $Y_{\alpha a}$, 
i.e., the right-handed neutrinos.
In terms of the Majorana fields
\be 
\chi = \chi_R + (\chi_R)^c \ , 
\qquad
N= N_R + (N_R)^c \ , 
\ee
the fermionic part of the Lagrangian (\ref{eq:lag1}) can be written as 
\be
\label{eq:lag2}
{\cal L} = 
- \frac{\lambda_\chi}{\sqrt{2}}  \left( \phi \, \overline{\chi} P_L \chi  + 
\phi^*  \, \overline{\chi} P_R \chi  \right)
-   \frac{\lambda_N}{\sqrt{2}} \left( \phi \, \overline{N} P_L N+ 
\phi^*  \, \overline{N} P_R  N   \right) - 
(Y \overline {L} H P_R N  + h.c.)
\ee
After the $U(1)_{B-L}$ symmetry breaking, the chiral fermions acquire Majorana masses, 
$m_\chi = \lambda_\chi v_\phi$, $m_N = \lambda_N v_\phi$, and ${\cal L}$
becomes
\bea
\label{eq:lag3}
{\cal L} &=& 
- \frac{m_\chi}{2} \overline{\chi} \chi   
- \frac{\lambda_\chi}{2}  \left[ (-h \, \sin \theta + \rho \cos \theta)  \overline{\chi} \chi  
- i \eta  \overline{\chi} \gamma_5 \chi  \right]
\\
&-& \frac{m_N}{2} \overline{N} N 
-   \frac{\lambda_N}{2} \left[  (-h \, \sin \theta + \rho \cos \theta)   \overline{N}  N  - i 
\eta  \overline{N} \gamma_5  N   \right] - 
(Y \overline {L} H P_R N + h.c.)  \nonumber 
\eea

Note that so far we have considered $N$ and $\chi$ as Majorana fields, yet degeneracies in the fermion mass matrix could lead to Dirac states. Indeed, one could find UV models where the structure of $\lambda_{ab}$ leads to two nearby states $\chi_1$ and $\chi_2$ which then would form a Dirac Dark Matter candidate~\cite{DeSimone:2010tf}. An example of this idea has been discussed in the previous section~\ref{sec:setup} under exotic lepton number.
See \cite{Racker:2014uga} for an alternative realization of the global $U(1)_{B-L}$ symmetry 
in which the dark fermions can naturally be pseudo-Dirac, in the context of an extended seesaw
scenario for neutrino masses.

\subsection{Neutrino masses}\label{sec:nmass}
In this section we briefly review the generation of (light) neutrino masses, namely TeV scale seesaw mechanism of type I. 
We denote $\nu_\alpha$ the active neutrinos and $N_s'$ the sterile ones. 
After electroweak symmetry breaking, the neutrino mass matrix in the basis 
$(\nu_\alpha, N_s')$ is given by
\be
\label{eq:numass}
{\cal M}_\nu = \left(
\begin{array}{cc}
0 & m_D \\
m_D^T & m_N
\end{array}
\right) \ , 
\ee
where $m_D = Y v_H/\sqrt{2}$ and $Y_{\alpha s}$ are the Yukawa couplings. Without loss of 
generality we can take the sterile neutrino Majorana mass matrix $m_N$ real and diagonal
in the $N'$ basis.

The matrix 
${\cal M}_\nu$ can be diagonalized by a unitary matrix $U$, so that 
\be
{\cal M}_\nu = U^* \, Diag(m_\nu,M) \, U^\dagger \ , 
\ee
where 
$m_\nu$ is the diagonal matrix with the three lightest eigenvalues of ${\cal M}_\nu$,  
of order $m_D^2/m_N$, 
and $M$ 
contains the heavier ones, of order $m_N$. 
 
The physical neutrinos ${\bf n}=(\nu_i,N_h$) are related to the active and sterile ones, 
($\nu_\alpha$, $N_{s}'$) by
\bea
\left(\begin{array}{c}\nu_\alpha \\ N_s' \end{array}\right)_L  =  U^* \,  
\left(\begin{array}{c}\nu_i  \\ N_h \end{array}\right)_L  \ .
\eea
The unitary matrix $U$ can be written as 
\be 
\label{eq:mixing}
U = \left(
\begin{array}{cc}
U_{\alpha i } & U_{\alpha h}  \\
U_{s i } & U_{s h } 
\end{array}
\right) \ , 
\ee
where, at leading order in the seesaw expansion parameter, ${\cal O}(m_D/m_N)$:

\bea
U_{\alpha i } &=& [U_{PMNS} ]_{\alpha i} \qquad   U_{sh} = I 
\nonumber \\
U_{\alpha h } &=&  [m_D m_N^{-1}]^*_{\alpha h}
\\
U_{s i} &= & - [m_N^{-1} m_D^T \, U_{PMNS}]_{si} \ .
\nonumber 
\eea
Notice that at this order the states $N$ and $N'$ coincide, so we identify them in the 
rest of this paper. 

Neglecting the mixing between the CP-even scalars, the Yukawa coupling of the SM-like  Higgs field  $h$  to the neutrinos can be written as 
\cite{Pilaftsis:1991ug}:
\be
{\cal L}_Y = - \frac{ h}{2 v_H}  \bar {\bf n}_i [ ( m_i + m_j) Re (C_{ij} )+ 
i \gamma_5 (m_j - m_i) Im (C_{ij}) ]  {\bf n}_j \ ,
\ee
where the indices $i,j$ refer to the light neutrinos $\nu_i$ 
for $i,j =1,2,3$ and to $N_h$ for $i,j =4,5,6$, and the matrix $C$ can be written in terms of
the mixing matrix $U$ as:
\be
\label{eq:cij}
C_{ij} = \sum_{\alpha=1}^{3} U_{\alpha i} U^*_{\alpha j} \ .
\ee

\section{Phenomenology}~\label{sec:pheno}
In this section we study the phenomenology of the proposed scenario. The main features of the model are determined by the interactions within the dark sector, i.e. Dark Matter, right-handed neutrinos, the scalar mediator $\rho$ and the Majoron $\eta$, and communication of the dark sector with the Higgs and leptons. This Table summarizes the source of constraints on the parameters of our model which we will explore in this section:
\bea
\begin{array}{|c|c|}  
\hline 
\textrm{ Parameter } & \textrm{ Constraint } \\\hline \hline 
\textrm{ Mixing } h \textrm{ and } \rho & {\rm BR_{inv}} \textrm{ and DD} \\\hline
\textrm{ Mixing } N \textrm{ and } \nu & h \to \textrm{exotic } \\\hline 
\textrm{ Dark } \chi \ ,  N \textrm{ and } \rho & \Omega_{DM} \textrm{, DD and ID}  \\\hline 
\textrm{Majoron } \eta  & N_{eff} \textrm{ and SIDM} \\\hline\end{array}
\nonumber
\eea

The mixing of the two scalars, the Higgs $h$ and $\rho$, is tightly constrained by {\it 1.)} limits on the Higgs invisible width ${\rm BR_{inv}}$ and global fits on Higgs properties as discussed on Sec.~\ref{sec:Hinv} and {\it 2.)} limits on direct detection (DD) from LUX~\cite{Akerib:2012ys,Akerib:2015rjg}, and XENON1T~\cite{Aprile:2015uzo} in the near future, see Sec.~\ref{sec:DD}. 
 
The mixing of the dark fermions $N$ and the left-handed SM neutrinos via their coupling to the Higgs produces a spectrum of massive neutrinos (see Sec.~\ref{sec:nmass}), but also leads to exotic decays of the Higgs to a dark fermion and a light neutrino. These are discussed in Sec.~\ref{sec:Hinv}. 

The interactions and masses of the dark fermions, Dark Matter $\chi$ and heavy neutrinos $N$, and the dark scalar $\rho$ can be probed in several ways. In sec.~\ref{sec:relic} we explain constraints from relic abundance $\Omega_{DM} h^2$ from Planck~\cite{Planck:2006aa,Ade:2015xua}, which provide information on the interplay among the competing annihilation processes, mainly the balance between the right-handed neutrino channel $\chi \chi \to N N$ and the annihilation to dark scalars, $\chi \chi \to \eta \rho$ and $\to \eta \eta$. Direct detection (DD) would provide complementary information, but it relies on the mixing of the dark scalar to the Higgs as mentioned above. Finally, annihilation of Dark Matter today could lead indirect detection (ID) signatures, namely features in the gamma-ray spectrum and signals in neutrino telescopes. These are discussed in Sec.~\ref{sec:ID}.

Finally, properties of the Majoron dark scalar $\eta$ can be probed by imprints in the CMB, such as $N_{eff}$ (see Sec.~\ref{sec:ID}) as well as by constraints on self-interacting Dark Matter (SIDM) which come from lensing and numerical simulations.

To deduce the constraints, we perform a simple Monte Carlo scan over the parameters in logarithmic scale, restricting the values of the couplings to the perturbative range, $\lambda_{\chi,N,\phi}\lesssim  {\cal O}(1)$ and the masses in the region of interest, $m_\chi$ \& $m_N$ from 1 GeV to 2 TeV, $m_\rho$ from 0.1 GeV to 10 TeV and $|\theta|$ from $10^{-4}$ to $\pi$. For the numerical implementation we made use of \code{LanHep}~\cite{Semenov:2008jy} and \code{micrOMEGAs}~\cite{Belanger:2013oya} in order to obtain the correct relic abundance, Higgs decays and today's annihilation cross section. We calculate $10^6$ points that match the {\it Planck} constraint on the Dark Matter abundance at $3\sigma$~\cite{Ade:2015xua}, namely $\Omega h^2 = 0.1198 \pm 0.0045$. 

\subsection{Constraints from Higgs decays }~\label{sec:Hinv}
\begin{figure}[t!]
\centering
\includegraphics[scale=0.15]{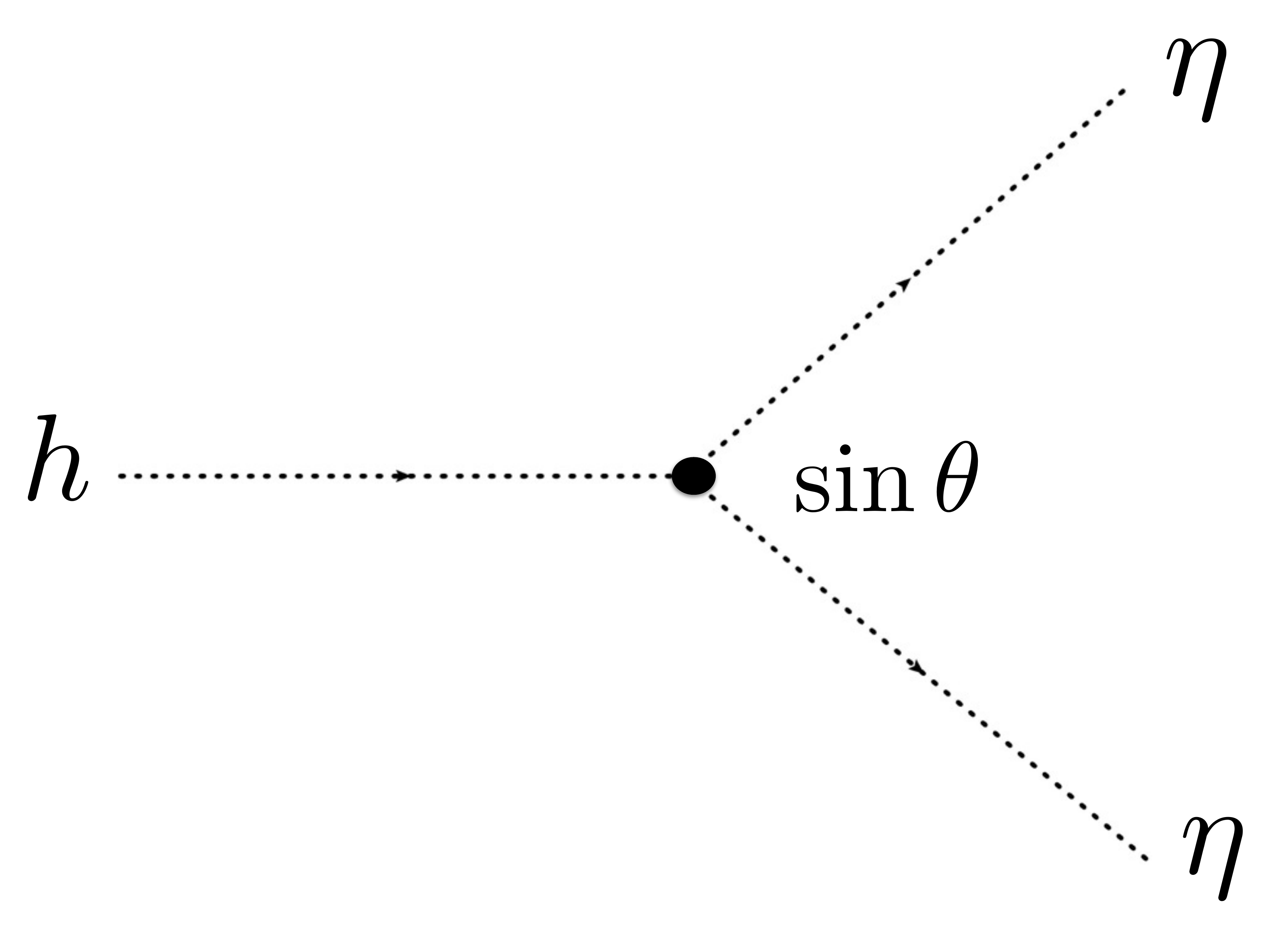}\vspace{1.cm}
\includegraphics[scale=0.15]{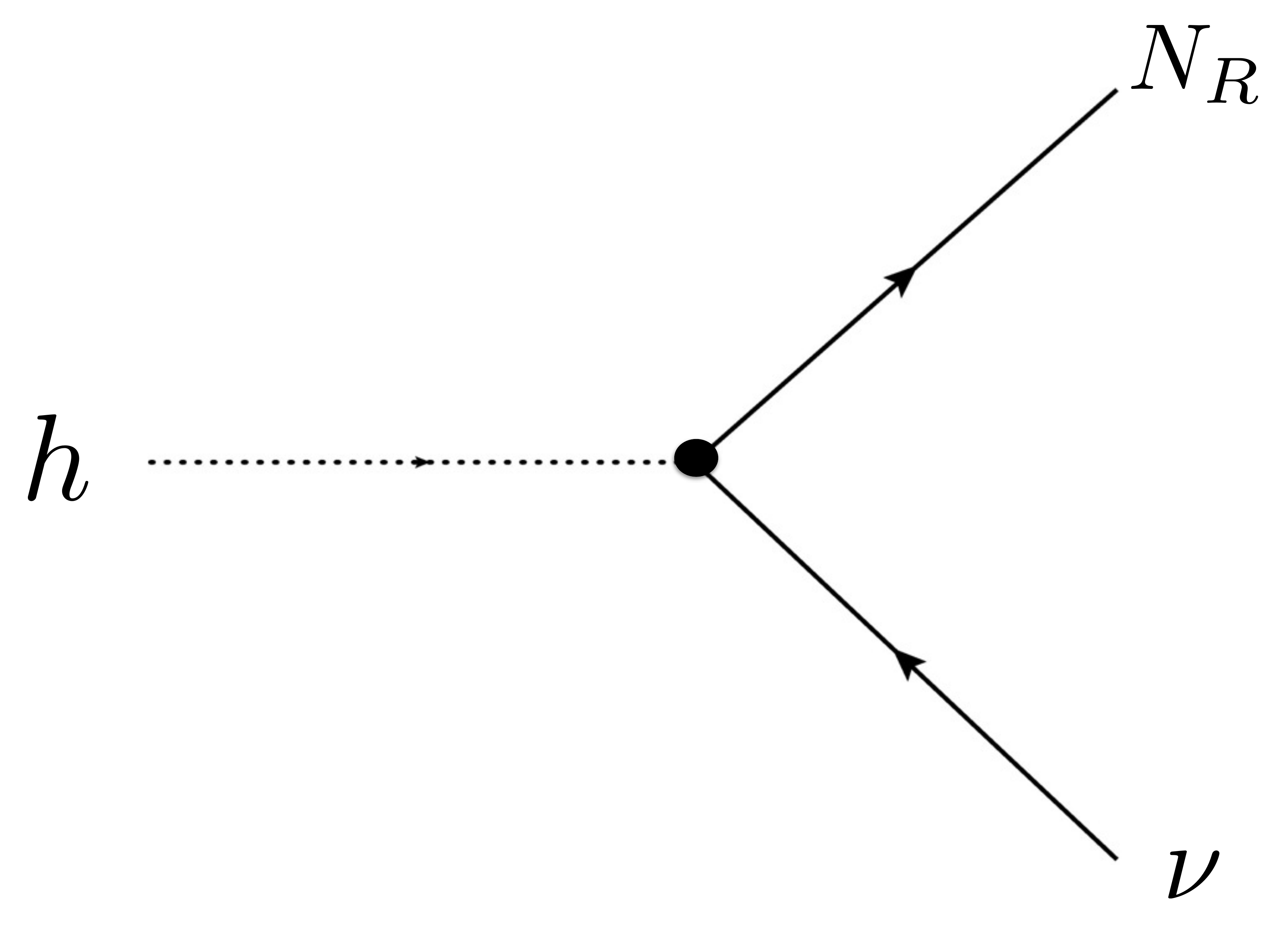}
\caption{(Left) Decay of the Higgs to two Majorons $\eta$ via the mixing of the Higgs with $\rho$. (Right) Exotic decay of the Higgs into a light neutrino and a dark fermion via their mixing.}\label{fig:hdec}
\end{figure}

In the two scenarios that we consider,
the enlarged fermion and scalar sectors lead to new decays of the Higgs boson, $h$ as shown in Fig.~\ref{fig:hdec}.
ATLAS and CMS constrain the invisible Higgs decay branching fraction as~\cite{ATLAS-CONF-2015-044,CMS-PAS-HIG-15-002} 
\be
\label{eq:h_inv}
{\rm BR_{inv}} = \frac{\Gamma_{\rm inv}}{\Gamma_{\rm inv} + \Gamma_{\rm SM}} 
< 0.23 \qquad (95 \% {\rm CL})\ ,
\ee 
where the SM Higgs width is $\Gamma_{\rm SM} \approx 4$ MeV.

The mixing of the two CP-even real scalars induce the following decay channels: 
\bea
\Gamma(h \rightarrow \eta \eta) &=& \frac{m_h^3}{32 \pi v_\phi^2} \sin^2 \theta 
\\
\Gamma(h \rightarrow \rho \rho) &=& \frac{(m_h^2 + 2 m_\rho^2)^2}{128 \pi m_h^2 v_H^2 v_\phi^2} \sqrt{m_h^2 - 4 m_\rho^2} (v_H \cos\theta - v_\phi \sin\theta )^2  \sin^2 2\theta 
\\
\Gamma(h \rightarrow \chi \chi) &=& \frac{\lambda_\chi^2}{16 \pi} 
\left(1 - \frac{4 m_\chi^2}{m_h^2} \right)^{3/2} m_h   \sin^2 \theta  \\
\Gamma(h \rightarrow N N ) &=&  \frac{\lambda_N^2}{16 \pi} 
\left(1 - \frac{4 m_N^2}{m_h^2} \right)^{3/2} m_h   \sin^2 \theta  ,
\eea
where we have neglected contributions to $h \to NN$ from the mixing among sterile and active neutrinos. This is justified by the smallness of the mixing, $\mathcal{O} (\sqrt{m_\nu/m_N})$. The decay to SM particles is modified as 
\be 
\Gamma (h \rightarrow {\rm SM \, particles}) =   \cos^2 \theta \, \Gamma_{\rm SM}  \ . 
\ee 
These global modifications of the Higgs couplings are equivalent to the well-studied case of mixing of the Higgs with a singlet and are well constrained~\cite{ATLAS-CONF-2015-044,CMS-PAS-HIG-15-002}. In the low $m_\rho$ region, the constraints one obtains from the invisible width is of the same order as this overall shift, hence below we use ${\rm BR_{inv}}$ as experimental input. Note that the corresponding expressions for $\rho$ decays widths are obtained by exchanging 
$\sin \theta \rightarrow \cos\theta$ and $m_h \rightarrow m_\rho$.

From the equation of the $h$ decay rate into two Majorons, $\Gamma(h \rightarrow \eta \eta)$, 
the experimental upper limit on the invisible decay width of the Higgs boson leads to the following upper bound on the mixing angle 
$\theta$ \cite{Weinberg:2013kea}: 
\be
|\tan \theta | \lesssim \sqrt{\frac{32 \pi v_\phi^2 \Gamma^{\rm SM}_{\rm Higgs}  {\rm BR_{inv}}}
{m_h^3 (1 - {\rm BR_{inv}})}} \sim 2.2 \times 10^{-3} \left(\frac{v_\phi}{10 \, {\rm GeV}} \right)
\ee
Including the other decay processes, when kinematically allowed, would reduce further the 
upper limit.

The Yukawa interaction term $Y \overline {L} H P_R N $ also leads to novel Higgs decay channels into neutrinos, 
even in the absence of mixing between de CP-even scalars. The corresponding decay width reads (for $\theta = 0$): 
\be
\label{eq:h_ninj} 
\Gamma(h \rightarrow  {\rm n}_i {\rm n}_j ) = \frac{\omega}{8 \pi m_h} 
\lambda^{1/2}(m_h^2,m_i^2,m_j^2) \left[
S \left(1 - \frac{(m_i+m_j)^2}{m_h^2} \right) 
+ P \left(1 - \frac{(m_i-m_j)^2}{m_h^2} \right) \right ] \ , 
\ee
where $\lambda(a,b,c)$ is the standard kinematic function, $w=1/n!$ for $n$ identical 
final particles and  
the scalar and pseudoscalar couplings are: 
\be 
S = \frac{1}{v_H^2} [ (m_i+m_j) Re(C_{ij})]^2  
\qquad , \qquad
P=  \frac{1}{v_H^2} [ (m_j- m_i) Im (C_{ij})]^2   \ , 
\ee
with $C_{ij}$  defined in eq.~(\ref{eq:cij}).
The largest branching ratio is for the decay into one light and one heavy neutrino
\cite{Gago:2015vma}:
\be 
\Gamma (h \rightarrow \nu N) = \frac{m_N^2}{8 \pi v_H^2}   
\left( 1 - \frac{m_N^2}{m_h^2} \right)^2 m_h | C_{\nu N }|^2  \ .
\ee
The attainable values for the above branching fractions have been analyzed in 
\cite{Gago:2015vma}, for the case of two heavy neutrinos, 
parameterizing the Yukawa couplings in terms of the observed light neutrino 
 masses and mixing angles, and a complex orthogonal matrix. 
 After imposing the relevant constraints from neutrinoless double beta decay, lepton flavour violating processes and direct searches of heavy neutrinos, they find that branching 
 ratios of $h \rightarrow \nu N $ larger than 
  $10^{-2}$ are generally ruled out for heavy neutrino masses 
 $m_N \leq 100$ GeV, and typically they are much smaller, due to the tiny Yukawa couplings
 required to fit light neutrino masses with sterile neutrinos at the electroweak scale.
 Therefore, the contribution of such decay modes to the Higgs decay width is 
 negligible, and they do not alter the bounds discussed above.

\subsection{Direct detection}~\label{sec:DD}
\begin{figure}[t!]
\centering
\includegraphics[scale=0.15]{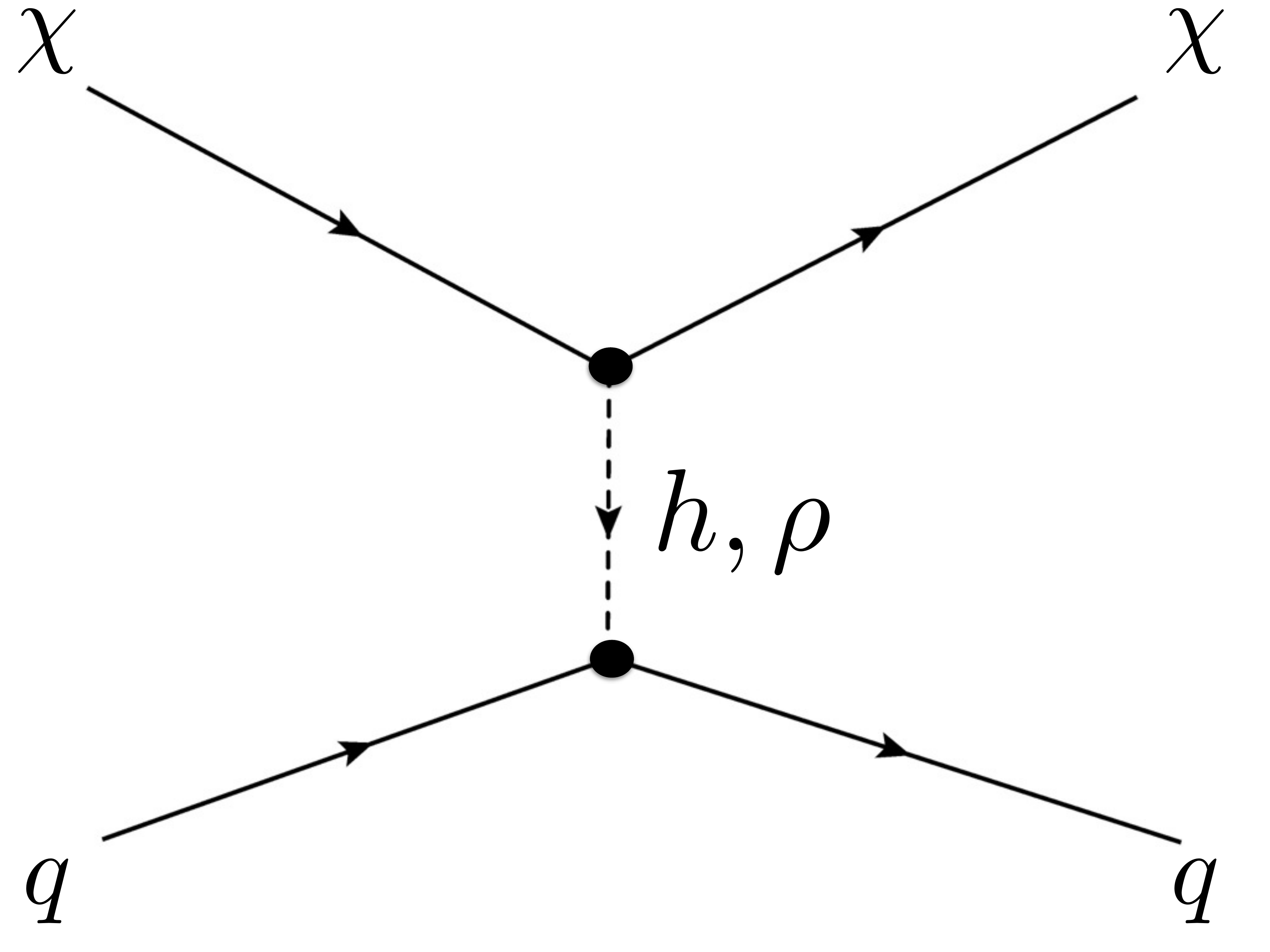}
\caption{Dark matter interaction relevant to direct detection constraints.}\label{fig:DD}
\end{figure}

In this scenario, Dark Matter scattering on nuclei relevant for 
direct Dark Matter detection is mediated via $t$-channel exchange by the CP even mass 
eigenstates, $h,\rho$ and it is spin-independent, see Fig.~\ref{fig:DD}.

The elastic scattering cross section of $\chi$ off a proton is given by \cite{Garcia-Cely:2013wda}:
\be
\label{eq:dd}
\sigma_{\chi p} = C^2 \, \frac{(\lambda_\chi \sin 2\theta)^2}{4 \pi v_H^2 } 
\frac{m_p^4 m_\chi^2}{(m_p+ m_\chi)^2} \left(\frac{1}{m_h^2} - \frac{1}{m_\rho^2} \right)^2 \ ,  
\ee
where $m_p$ stands for the proton mass and $C \simeq 0.27$ is a constant that depends on the nuclear matrix element \cite{Belanger:2013oya}. 

The constraints on the combination $\lambda_\chi |\sin 2\theta|$ both from the invisible 
Higgs decay width and from the LUX experiment \cite{Akerib:2013tjd}
have been thoroughly analyzed in \cite{Garcia-Cely:2013wda}, in a model where the 
Dark Matter is a chiral fermion charged under a global $U(1)$ symmetry spontaneously broken by a SM singlet scalar, $\phi$.
Although such scenario does not include the further interaction among $\phi$ and the
sterile neutrinos which can be present when the global $U(1)$ is identified with $U(1)_{B-L}$,
the limits from direct Dark Matter searches apply exactly the same, and the bounds from 
the Higgs invisible width will be even stronger in our case, since there are new non-standard 
decays contributing to it, namely $\Gamma(h \rightarrow N N )$. 

\begin{figure}[t!]
\begin{tabular}{c}
\includegraphics[width=0.98\textwidth]{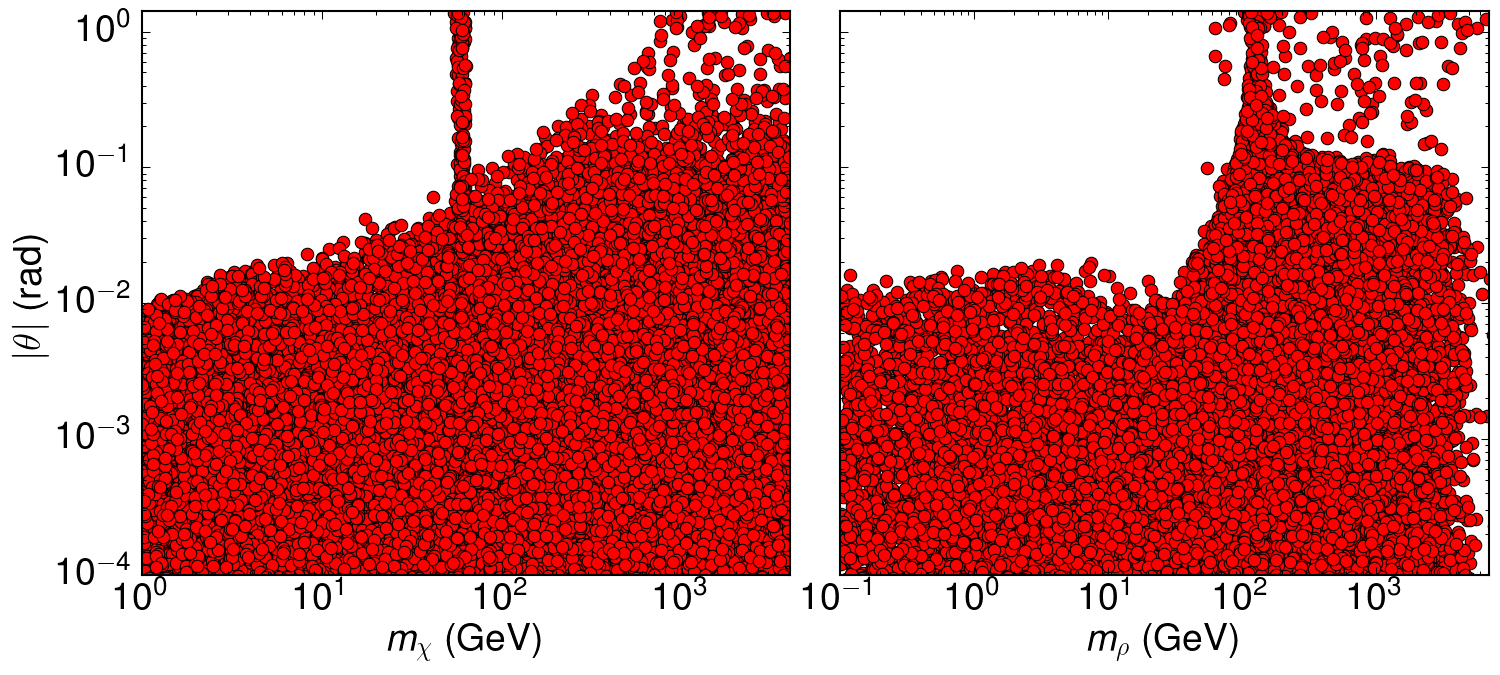} \\
\end{tabular}
 \caption{Allowed values of the mixing between the Higgs and the dark sector as a function of the Dark Matter mass (left) and the scalar mediator $\rho$ (right) after imposing the constraints from LUX and the Higgs invisible decay width. Large values of the mixing are possible only in the somewhat tuned regions $m_\chi \simeq m_h/2$ and $m_\rho \simeq m_h$.}\label{fig:DD_miguel}
\end{figure}

We have analyzed the constraints on the mixing in our model, and found that for low values of $m_\chi$ and $m_\rho$ the stronger limit comes from the invisible Higgs decay width,
while for higher masses the bound is determined by direct detection experiments. When applying both constraints altogether they exclude $\theta \gtrsim 0.1$ for all parameter space but the region $m_\rho \simeq m_h$, where a cancellation in the direct detection cross section occurs, see eq.~(\ref{eq:dd}), and the regions $m_\chi \simeq m_h/2$,
$m_\chi \simeq m_\rho/2$, where a resonance in the Dark Matter annihilation cross section mediated by the Higgs  or the $\rho$ occurs.
These results are shown in Fig.~\ref{fig:DD_miguel}, where the allowed values of the mixing as a function of the Dark Matter mass and mediator $\rho$ are shown back-to-back, to illustrate the correlation between the points of $\theta \gtrsim 0.1$ and the regions where $m_\chi \simeq m_h/2$ (left) or $m_\rho\simeq m_h$ (right).

\subsection{Dark Matter relic abundance}\label{sec:relic}
\begin{figure}[t!]
\begin{center}
\begin{tabular}{c}
p-wave \\
\end{tabular}

\begin{tabular}{cc}
\includegraphics[width=0.3\textwidth]{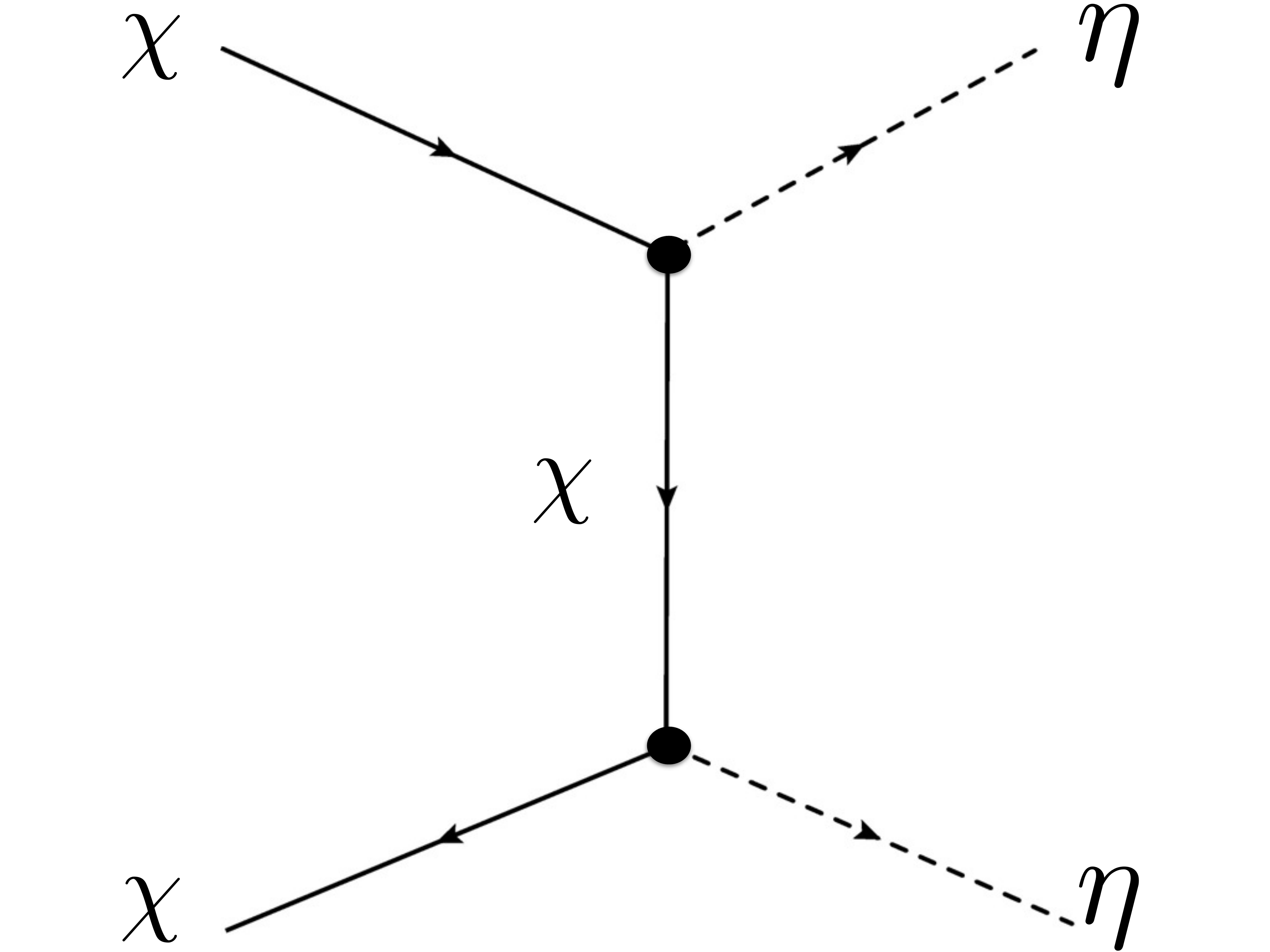} & \includegraphics[width=0.3\textwidth]{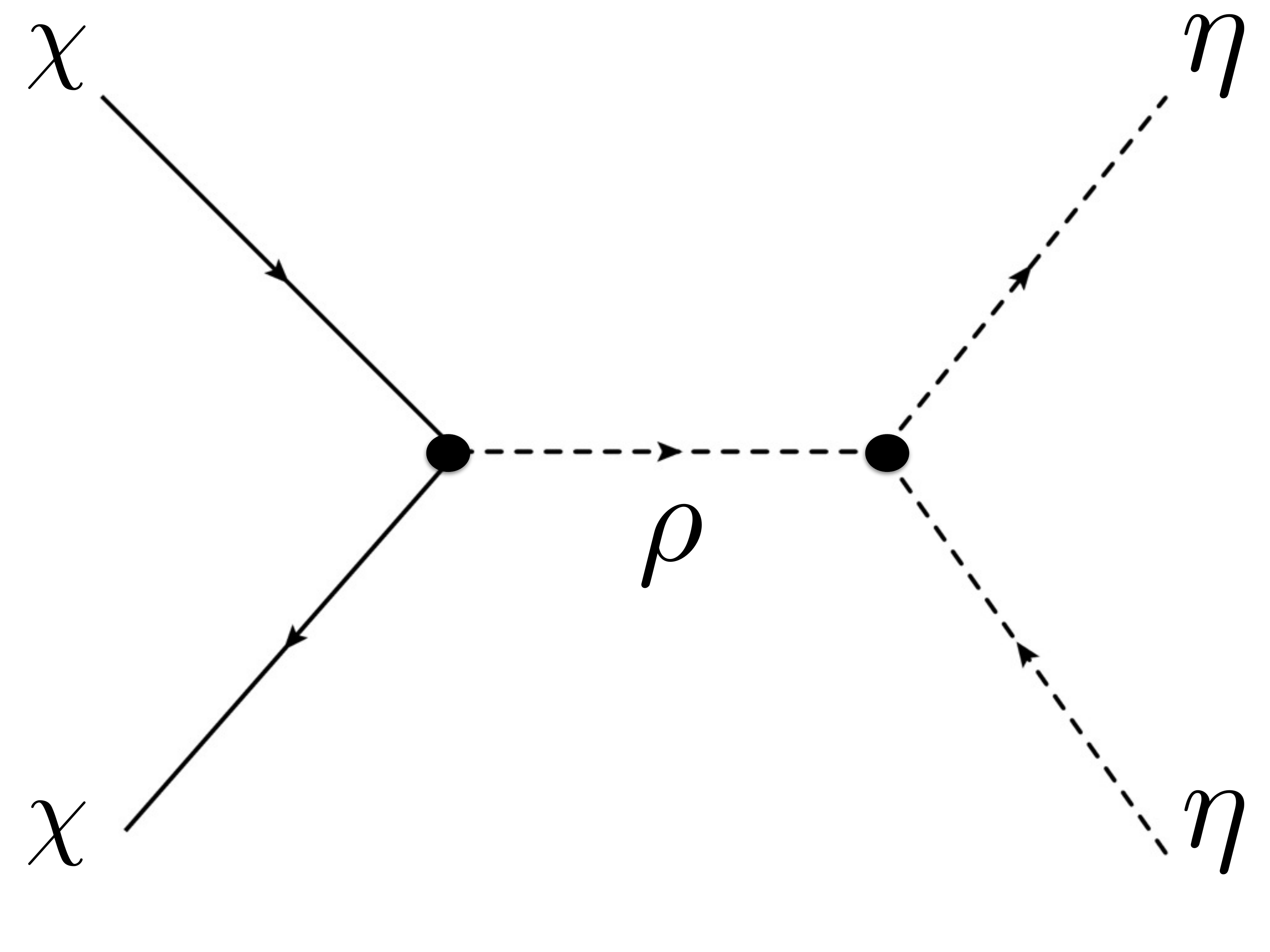}\\
\end{tabular}
\begin{tabular}{ccc}
s-wave & s-wave & s-wave $m_N^2$ suppressed \\
\includegraphics[width=0.3\textwidth]{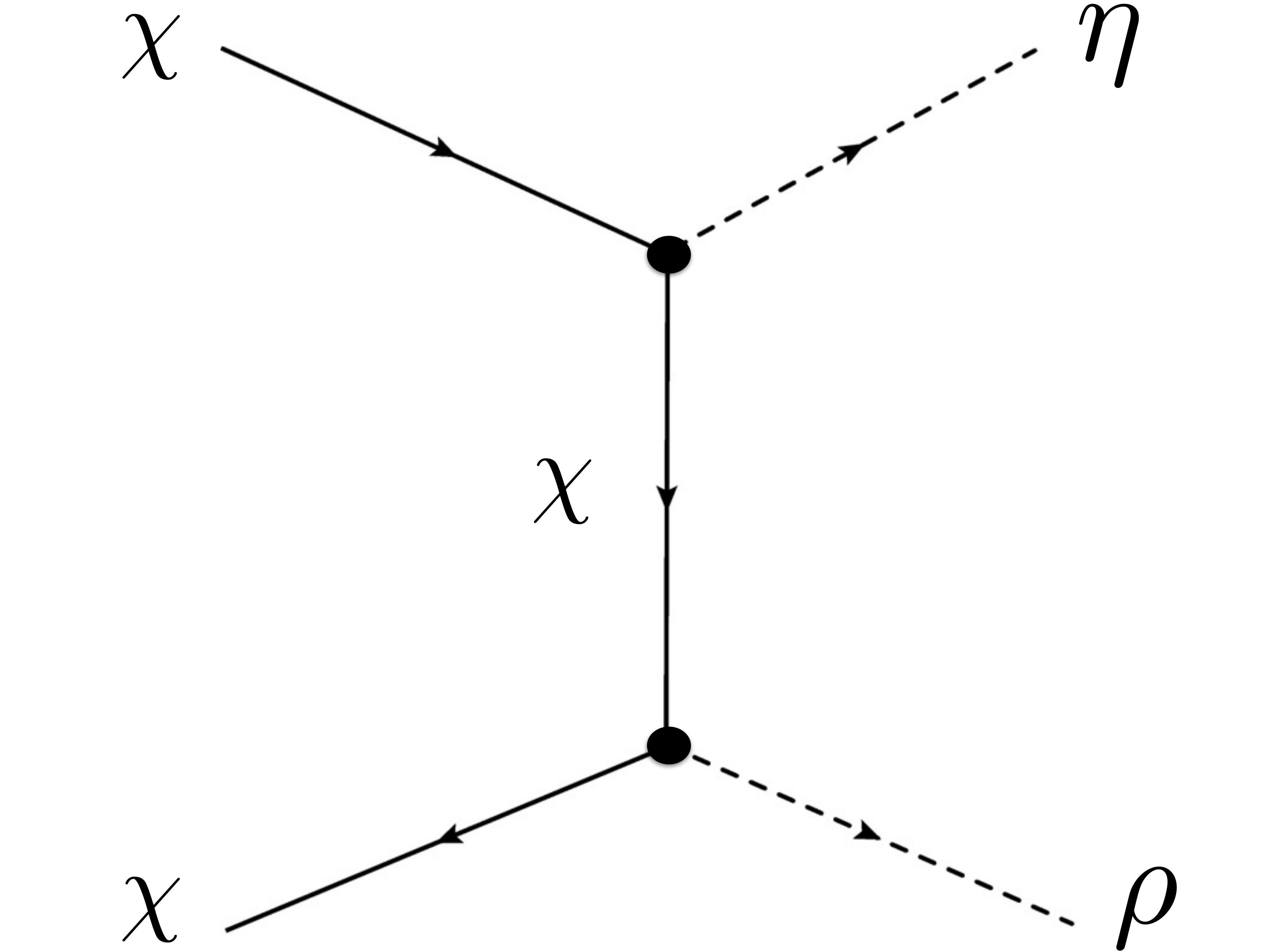} &
\includegraphics[width=0.3\textwidth]{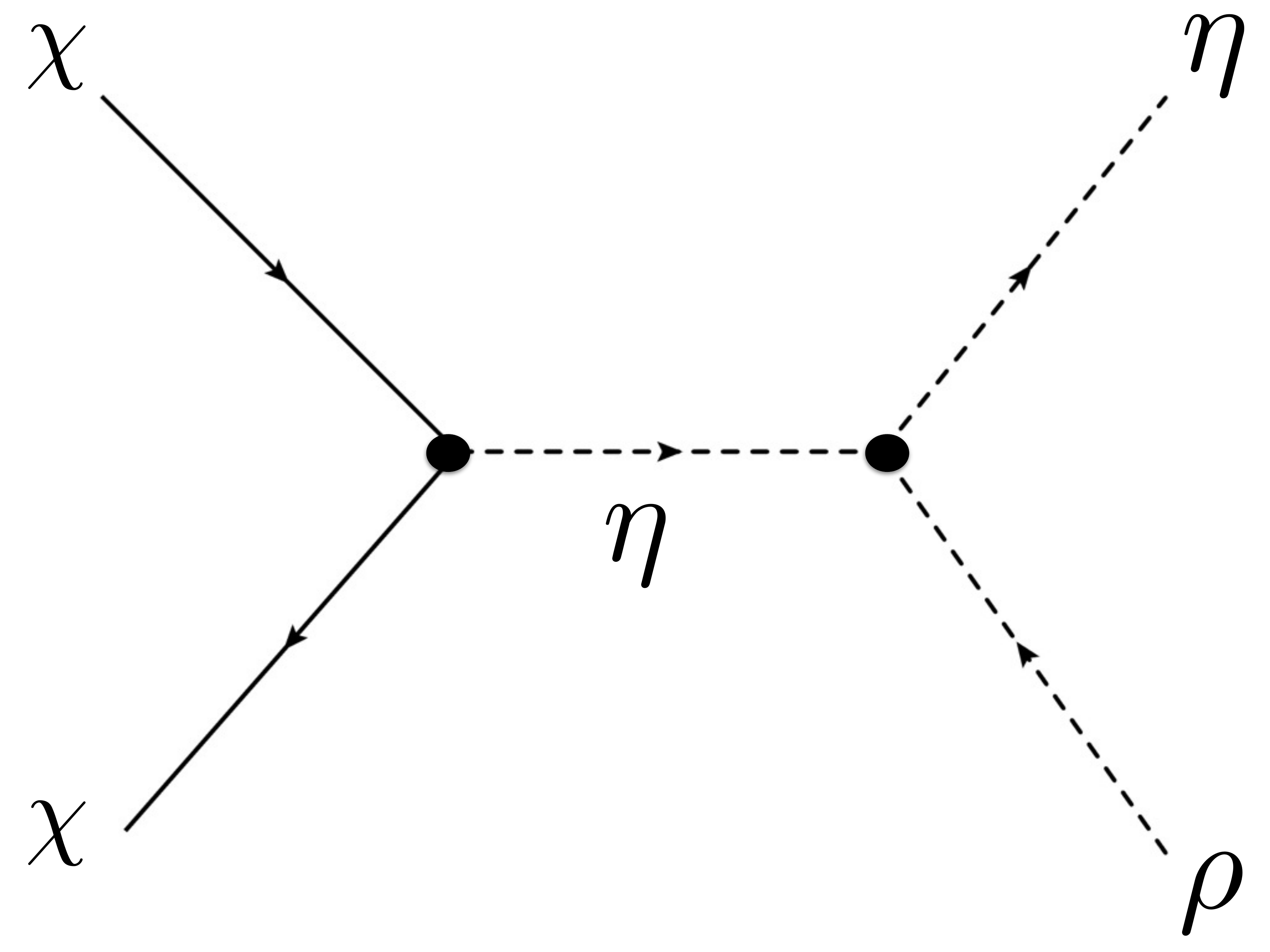} &
\includegraphics[width=0.3\textwidth]{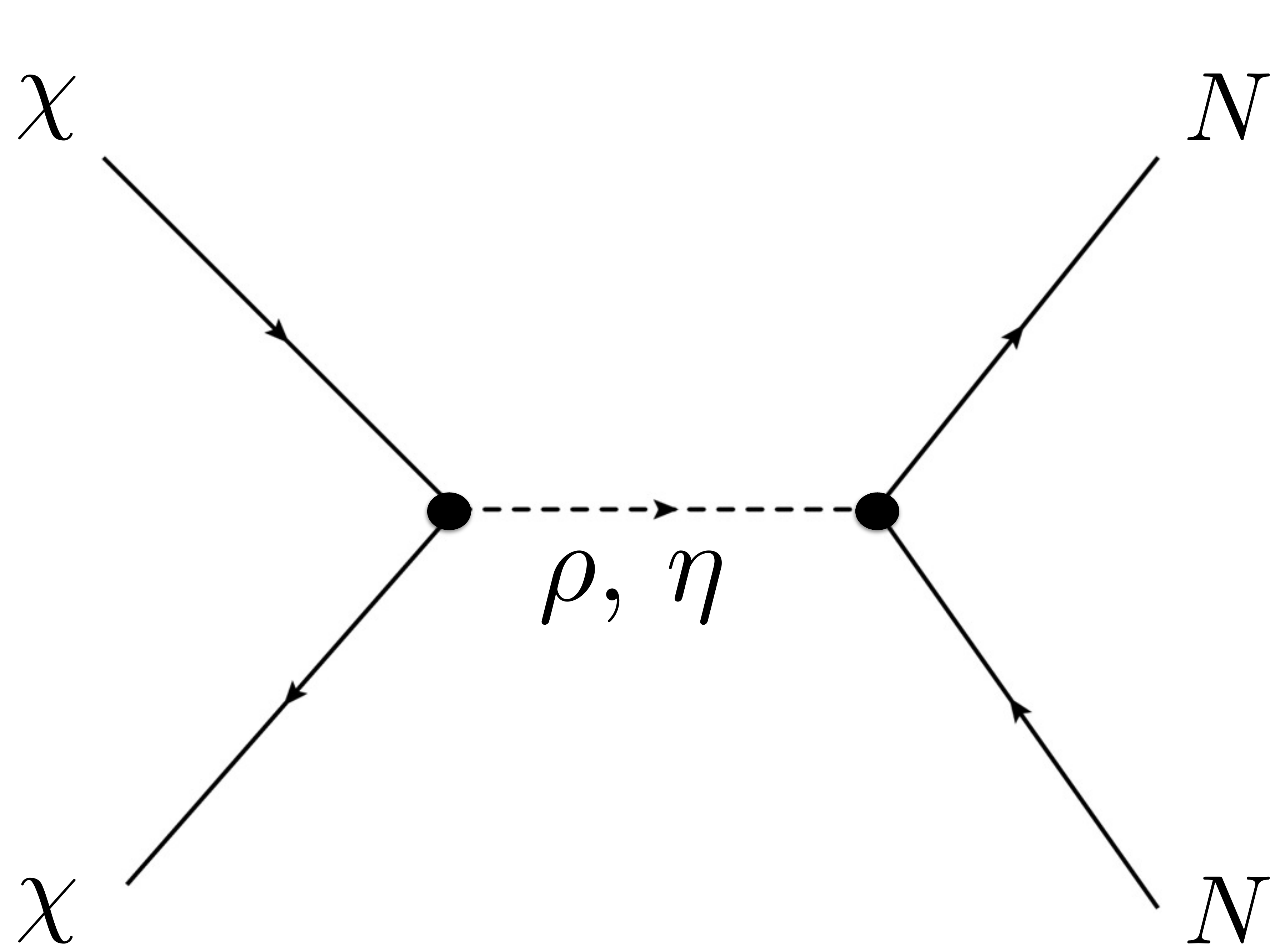} \\
\end{tabular}

\end{center}
 \caption{Diagrams relevant to the relic abundance computation.}
\label{fig:chichi_relic}
\end{figure}

The Dark Matter annihilations into SM particles are strongly suppressed
due to the bounds on $\theta$ discussed in the previous section,
except when $m_\rho \simeq m_h$ and in the resonance regions $m_\chi \sim m_h/2$ or 
$m_\chi \sim m_\rho/2$. Moreover such annihilations channels are p-wave suppressed. 
 Keeping in mind that these somehow fine-tuned possibilities are always 
open, we focus on the dominant annihilation channels, 
involving the new scalars $\rho,\eta$ as well as the sterile neutrinos $N$, see Fig.~\ref{fig:chichi_relic}. 
For simplicity, in the following we will only consider one generation of right-handed neutrinos, but extending the discussion to more generations will be straightforward.
Therefore, in order to reduce the large number of free parameters
in this analysis we set the mixing angle $\theta=0$, 
and we scan over the remaining independent variables, chosen to 
be $m_\chi, m_N, m_\rho$ and $\lambda_\chi$.

There are two channels with s-wave annihilation cross-section, the production of two right-handed neutrinos and the final state $\eta \rho$,
\begin{eqnarray}
\sigma_{\chi\chi\to \rho \eta} v_{rel} 
&=&  \frac{m_\chi^2}{1024 \pi  v_\phi^4}  \,  \left(4 -r_\rho^2\right)^3
 \, + {\cal O}(v_{rel}^2)  \ , \nonumber \\
\sigma_{\chi\chi\to NN} v_{rel} 
&=& \frac{m_N^2 }{64 \pi  v_\phi^4} \, \sqrt{1 -r_N^2}
 \, + {\cal O}(v_{rel}^2)  \ .
\end{eqnarray}

Other possible channels are p-wave suppressed,
\begin{eqnarray}\label{fig:rhorho}
\sigma_{\chi\chi\to \eta \eta} v_{rel}
& =& \frac{m_\chi^2}{192 \pi  v_\phi^4} \, \frac{8 +r_\rho^4}{ \left(r_\rho^2-4\right)^2} \, v_{rel}^2  \ , \nonumber \\
\sigma_{\chi\chi\to \rho \rho}  v_{rel}
&=& \frac{m_\chi^2  }{384 \pi  v_\phi^4} \,   \frac{\sqrt{1-r_\rho^2}}{\left(r_\rho^2-4  \right)^2} \, 
\left( 144 - 32 r_\rho^2 \right)  \, v_{rel}^2 \  + {\cal O}(r_\rho^4).
\end{eqnarray} 

Here $v_{rel} = 2 \sqrt{1-4m_\chi^2/s}$ is the relative velocity of the Dark Matter in the 
center of mass frame and the ratios are given by $r_\rho=m_\rho/m_\chi$ and $r_N=m_N/m_\chi$.

\begin{figure}[t!]
\begin{center}
\includegraphics[width=0.98\textwidth]{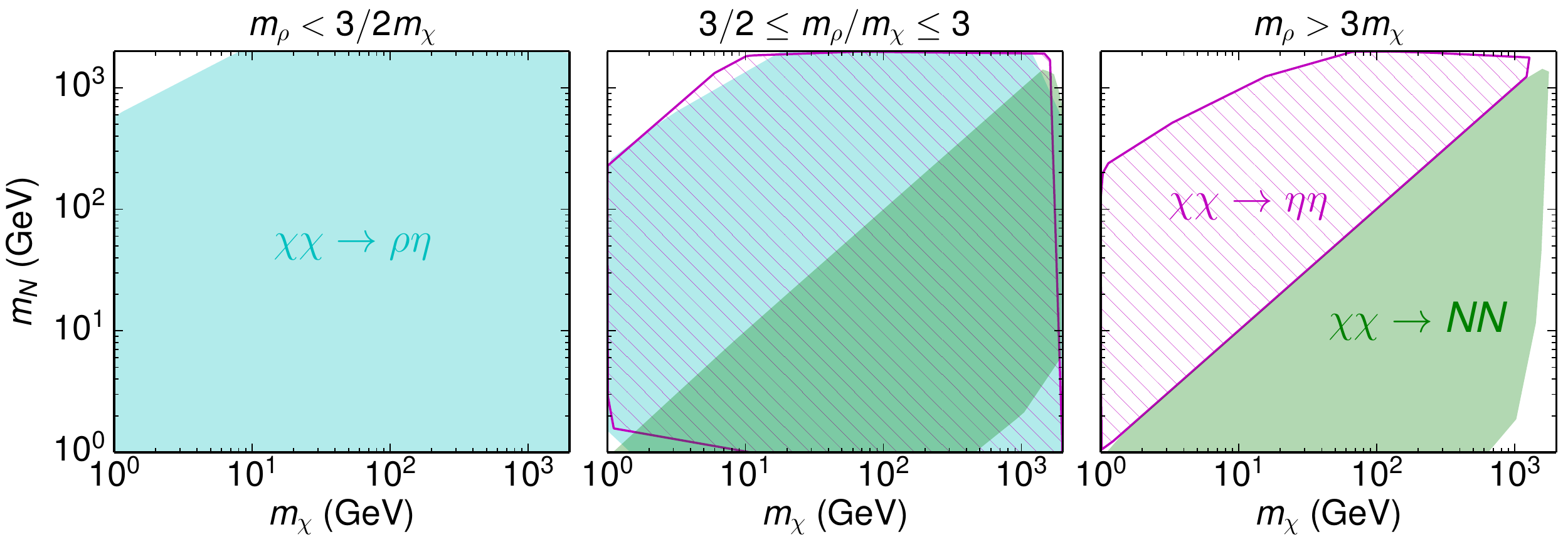} 
\end{center}
 \caption{Allowed values of dark fermion masses, $m_{\chi}$ and $m_N$. The different colours  correspond to regions in parameter space in which the annihilation channel constitutes more than 60\% of the total cross section for $v=10^{-3}c$,  as relevant for indirect detection.}
\label{fig:m_Nvsmchi}
\end{figure}

As the annihilation channel into sterile neutrinos is not velocity suppressed, it can be comparable to the scalar channels, 
$\chi \chi \rightarrow \eta \eta, \rho \rho, \eta \rho$, which alike the $NN$ channel, are not vanishing even in the case of zero $h-\rho$ mixing. 
We find that there is a significant fraction of the parameter space of the model in which the annihilation channel into $N N$ is relevant, and even dominant. This is shown in Fig.~\ref{fig:m_Nvsmchi}, where allowed values of dark fermion masses, $m_{\chi}$ and $m_N$ are depicted, with different colours corresponding to regions with dominance of one channel in the annihilation cross section for $v = 10^{-3}c$,  as relevant for the calculation of direct detection constraints. 
The three panels of Fig.~\ref{fig:m_Nvsmchi} correspond to different ranges of the dark scalar mass, $m_\rho$. In particular, in the middle panel we have singled out the region $3/2 \leq m_\rho / m_\chi \leq 3 $, where the annihilation into $\eta \eta$ is resonantly enhanced.

On the other hand, the annihilation channel $\eta  \rho$ tends to dominate when kinematically accessible (i.e., in the region $m_\rho < 2 m_\chi$, since we are neglecting the mass of the Majoron, $m_\eta$), as it is parametrically enhanced respect to the other s-wave channel into right-handed neutrinos by $(m_\chi/m_N)^2$.
Finally, in the range $m_\rho > 2 m_\chi$, the two channels 
$N N $  and $\eta \eta$ compete:
the former dominates when 
 $m_N \lesssim m_\chi$ and $m_\rho > 3 m_\chi$, while 
in the resonance region $3/2 \leq m_\rho / m_\chi \leq 3 $  we find that any of the two annihilation channels may dominate, as well as the $\eta \rho$ if open (central panel).

Interestingly,  the $\eta \eta$ channel could have
 dominated the dynamics at freeze out, with the $NN$ channel playing an spectator role, but for the usual velocities that the Dark Matter particles have in the galactic halo $v \simeq 10^{-3}c$,  the $NN$ channel could dominate the Dark Matter annihilation at later times. However, even if dominant,  the cross section may be too small to lead to  any indirect detection signature (see Fig.~\ref{fig:s_ann}).


 Notice that due to the $U(1)_{B-L}$ symmetry,  there is a non trivial relation between the mass of the sterile neutrinos and the 
 Dark Matter annihilation cross section into them, since both are proportional to the coupling 
 $\lambda_N$. As a consequence, when the sterile neutrinos are very light, and the phase space 
 is more favourable, the coupling is too small and the annihilation into sterile neutrinos is 
 suppressed. On the other hand, if the coupling  $\lambda_N$ is large, the sterile neutrinos are too 
 heavy, and this annihilation channel is phase space suppressed or even forbidden.
 Due to this relation between sterile neutrino masses and coupling to the scalar $\phi$, the Dark Matter annihilation cross section into $N N $,  although s-wave, does not always 
 dominate over the $p$-wave suppressed $\eta \eta$ channel, 
 which can be enhanced  
 near the resonance, $m_\chi \sim m_\rho/2$. 
   
\subsection{Constraints from indirect searches and CMB}~\label{sec:ID}
Dark Matter is also searched for indirectly, through the detection of SM products (photons, neutrinos 
and antiparticles) from its annihilation (or decay) in dense regions of the Universe, such as the 
center of the Milky Way. In particular, detection of gamma rays and neutrinos are useful because the signal can be traced back to the source.  

Our scenario predicts that Dark Matter particles could be annihilating in the center of the galaxy through the s-wave processes $\chi \chi \rightarrow \eta\rho, N N $, if kinematically allowed. 
When the first annihilation channel dominates
there are not photons in the final state 
since $\rho$ also decays invisibly to $\eta \eta$. 
Although quantum gravity effects could explicitly break the global $U(1)_{B-L}$ symmetry providing a mass to the Majoron, 
it would decay into light neutrinos~\cite{Lattanzi:2014mia}, for which indirect detection constraints are quite weak, as we shall discuss below.
Therefore, we do not expect gamma ray signals from Majoron decays, and we are left with the constraints from the $NN$ channel.

Thus, in order to analyze the possible indirect detection signals we  
next discuss the relevant decay modes of the heavy neutrinos. 
In this section we also neglect the CP-even scalar mixing angle, $\theta$, which is 
tightly constrained by the invisible Higgs decay width and direct Dark Matter searches, 
so it will not affect our results below. 

  The decay channel $N \rightarrow \nu \eta$ is always open (provided the Majoron mass is 
  $m_\eta < m_N$),  
  and neglecting the masses of the decay products its partial width is given by \cite{GonzalezGarcia:1990fb}:
 \be
 \label{eq:nueta}
 \Gamma(N  \rightarrow \nu_j \eta) = \frac {m_{N}^3}{128 \pi v_\phi^2} \sum_j |R_{Nj}|^2  \ , 
 \ee
 where in the seesaw limit 
 $R_{Nj} = 2 U_{Nj} = - 2 (m_N^{-1} m_D^T U_{PMNS})_{Nj}$ (we have summed 
 over all the light neutrinos in the final state) and induces typical decay widths of the order 
 \bea
\Gamma(N  \rightarrow \nu \eta) \simeq \frac{1}{32 \pi} \, \left( \frac{m_N}{v_\phi}\right)^2 \, m_\nu \ ,
 \eea
so that they safely decay before Big-Bang nucleosynthesis.

  Regarding the decays to SM particles, they depend on the mass of the heavy lepton,
 $m_N$. In the following we discuss two cases, depending on whether the neutrino is heavier or lighter than massive vector bosons.
 
{\bf Light right-handed neutrino, $m_N < m_W$: } If the right-handed neutrino is lighter than the $W$ boson, $N$ will decay through off-shell $h,Z,W$
 bosons to three fermions. Since the decay via a virtual $h$ is further suppressed by the small Yukawa couplings of the SM fermions, it is a very good approximation to consider only the processes mediated by virtual $W,Z$, whose partial widths read~\cite{GonzalezGarcia:1990fb}:
\bea 
 \Gamma(N  \rightarrow \nu q \bar{q} ) &=& 3\, A C_{NN}
[2(a_u^2 + b_u^2) + 3(a_d^2 + b_d^2)] f(z) 
\\
 \Gamma(N   \rightarrow  3 \nu) &=& A C_{NN}
[\frac 3 4 f(z)  +  \frac 1 4  g(z,z)]  
\\
 \Gamma(N   \rightarrow \ell q \bar{q} ) &=& 6\, A C_{NN}
 f(w,0) 
\\
 \Gamma(N   \rightarrow \nu \ell\bar{\ell} ) &=&  A C_{NN}
[3(a_e^2 + b_e^2) f(z)  + 3 f(w) - 2 a_e g(z,w) ]
\eea
where $C_{NN}$ is defined in eq.~(\ref{eq:cij}), 
\be
A \equiv  \frac {G_F^2 m_{N}^5 }{192 \, \pi^3}  \ ,
\ee
$a_f,b_f$ are the left and right neutral current couplings of the fermions ($f=q,\ell$),  
the variables $z,w$ are given by 
\be 
z= (m_N/m_Z)^2  \ , \qquad w = (m_N/m_W)^2 \ , 
\ee 
 and the functions $f(z), f(w,0)$ and $g(z,w)$ can be found in \cite{Dittmar:1989yg}.

  Assuming no strong cancellations in the mixing matrix $U$, 
  we expect $C_{NN} \sim m_\nu/m_N$, so from the equations above we can estimate the ratio between the total decay width to three SM particles and the invisible decay width to $\nu \eta$,   given by eqs.~(\ref{eq:nueta}):
 \be
 \frac{\Gamma (N \rightarrow {\rm 3 \, SM})}{\Gamma(N   \rightarrow \nu \eta)} 
 \sim \frac{1}{\pi^2}  \left(\frac {m_N}{v_H} \right)^2  \left(\frac {v_\phi}{v_H} \right)^2  
 \lesssim 10^{-2} \left(\frac {v_\phi}{v_H} \right)^2  \ , 
 \ee
where in the last term we have used that  $m_N < 80$ GeV.
 Therefore the three-body decays to SM particles are suppressed when $m_N < m_W$ and the right-handed neutrino decays invisibly to $\nu \eta$, unless $v_\phi \gtrsim 10 \, v_H$.
 On the other hand,  
 the coupling between the sterile neutrinos and the Majoron $\eta$ is 
  $\lambda_N = m_N/v_\phi \lesssim 0.05 $ for $v_\phi \gtrsim$ 1.6 TeV, probably too small to have a significant 
  DM annihilation cross section into $ N N $ in the first place.

 Moreover, in the $N N$ annihilation channel also light neutrinos are copiously produced, which could lead to observable signals at IceCUBE. 
 These will depend on the neutrino energy, and therefore a detailed study of the final state spectrum is required to set constraints. 
Very roughly, for heavy Dark Matter we expect very energetic neutrinos, so that this scenario could be tested with current IceCUBE data, provided $E_\nu \gtrsim$ 100 GeV.
If the Dark Matter is lighter, or the neutrino energy spectrum softer, 
DeepCore will be needed to further constrain the parameter space, 
since it is expected to lower the IceCUBE neutrino energy threshold to about 10 GeV. 
 
{\bf Heavy right-handed neutrino, $m_N  > m_W$: } For larger values of $m_{N}$, two body decays to SM particles are open, and the corresponding widths 
 read \cite{Pilaftsis:1991ug}:
  \bea 
  \label{eq:WZh}
  \Gamma(N  \rightarrow W^{\pm} \ell^{\mp}_{\alpha}  ) &=& \frac{g^2}{64 \pi} |U_{\alpha N}|^2 
  \frac{m_N^3}{m_W^2} \left(1 -   \frac{m_W^2}{m_N^2} \right)^2
   \left(1 +  \frac{2 m_W^2}{m_N^2} \right)  
   \\
    \Gamma(N  \rightarrow Z  \, \nu_\alpha ) &=& \frac{g^2}{64 \pi c_W^2 } |C_{\alpha N}|^2 
  \frac{m_N^3}{m_Z^2} \left(1 -   \frac{m_Z^2}{m_N^2} \right)^2
   \left(1 +  \frac{2 m_Z^2}{m_N^2} \right)    
   \\
   \Gamma(N  \rightarrow h \,  \nu_\alpha ) &=& \frac{g^2}{64 \pi } |C_{\alpha N}|^2 
  \frac{m_N^3}{m_W^2} \left(1 -   \frac{m_h^2}{m_N^2} \right)^2  
  \eea
 
 In the above expressions, we have assumed that $N$ is a Majorana fermion.
 
 From eqs.~(\ref{eq:nueta}) and (\ref{eq:WZh}), we se that in this mass range the ratio between Majoron and SM particles decay widths is approximately given by
 \be
 \frac{\Gamma (N \rightarrow {\rm SM})}{\Gamma(N   \rightarrow \nu \eta)} 
 \sim \left(\frac {v_\phi}{v_H} \right)^2  \ .
 \ee

 Thus in this mass range we expect a significant flux of gamma rays from the galactic center produced by SM annihilation products, and bounds can be set from the Fermi-LAT Space Telescope gamma ray data.
A detailed study of the indirect detection signatures of our scenario is beyond the scope of this work, since Dark Matter does not decay directly to SM particles, as it is usually assumed in most analysis, but to two right-handed neutrinos that subsequently decay to them. 
 Therefore we just estimate here the expected constraints using current analysis. See Sec.~\ref{sec:results} for a discussion on how these limits affect the allowed region in the parameter space of our model.

Dark matter particles in the galactic halo can scatter elastically with a nucleus and become trapped in the gravitational well of an astronomical object, such as the Sun. They will undergo subsequent scatterings, and eventually thermalize and concentrate at the core of the object. The Dark Matter accumulated 
in this way may annihilate into SM particles, in particular neutrinos that can be detected by neutrino experiments
like IceCUBE or SuperKamiokande. However we do not discuss this type of indirect detection constraints here, since 
in our scenario the limits from direct searches are tighter and moreover
they can 
always be avoided with a small enough mixing angle between the CP-even scalars, which suppresses the DM-nucleon elastic cross-section still getting the 
correct Dark Matter relic abundance through annihilation into $N N $ or $\eta \rho$, which is our case.

Measurements of the cosmic microwave background (CMB) anisotropies are also sensitive to Dark Matter annihilation during the cosmic dark ages, because the injection of ionizing particles will increase the residual ionization fraction, broadening the last scattering surface and modifying the anisotropies.
Under the assumption that the power deposited to the gas is directly proportional to that injected at the same redshift, with some efficiency factor $f_{eff}$, constraints can be placed on the combination $f_{eff} \langle \sigma v \rangle/m_{DM}$, for different SM annihilation channels in
s wave.  
  Again, the available calculations of $f_{eff}$ assume that Dark Matter annihilates directly to a pair
  of SM particles, and thus they are not directly applicable to our model, but we can roughly estimate 
  the expected impact of such limits in the allowed parameter space assuming as before that the constraints will be similar for cascade decays.
  In \cite{Slatyer:2015jla}, $f_{eff}$ has been calculated as a function of the Dark Matter mass for a range of SM final states, and using the most recent results from the Planck satellite she found that for any linear combination of SM final states which does not contain a significant branching ratio of Dark Matter  annihilation directly into neutrinos one must have $\langle \sigma v \rangle \lesssim  3 \times 10^{-27} (m_{DM}/ 1 \, {\rm GeV) \, cm^3/s} $. 
 However in our scenario when the Dark Matter (and thus the sterile neutrino) 
 is lighter than $m_W$, the final states are $\nu, \eta$ and therefore the above limit does not apply.
 
  Only for higher Dark Matter masses the final annihilation products can be charged leptons and gauge bosons, 
  but in this range the CMB limits are above the thermal relic cross section, so they do not constrain our
 scenario.

The CMB also constrains the properties of the (massless) Majoron $\eta$, which constitutes a form of dark radiation and contribute to $N_{eff}$~\cite{Weinberg:2013kea}. In Ref.~\cite{Garcia-Cely:2013wda}, it was shown that typically the limit on $N_{eff}$ is already saturated by constraints from direct detection. In particular, for $m_\chi >$ 100 GeV, a non negligible contribution to dark radiation could only happen in a small range of $m_\rho$ from 0.5 to 1 GeV.        
   
\subsection{Self-interacting Dark Matter}\label{sec:SIDM}
\begin{figure}[t!]
\centering
\includegraphics[scale=0.2]{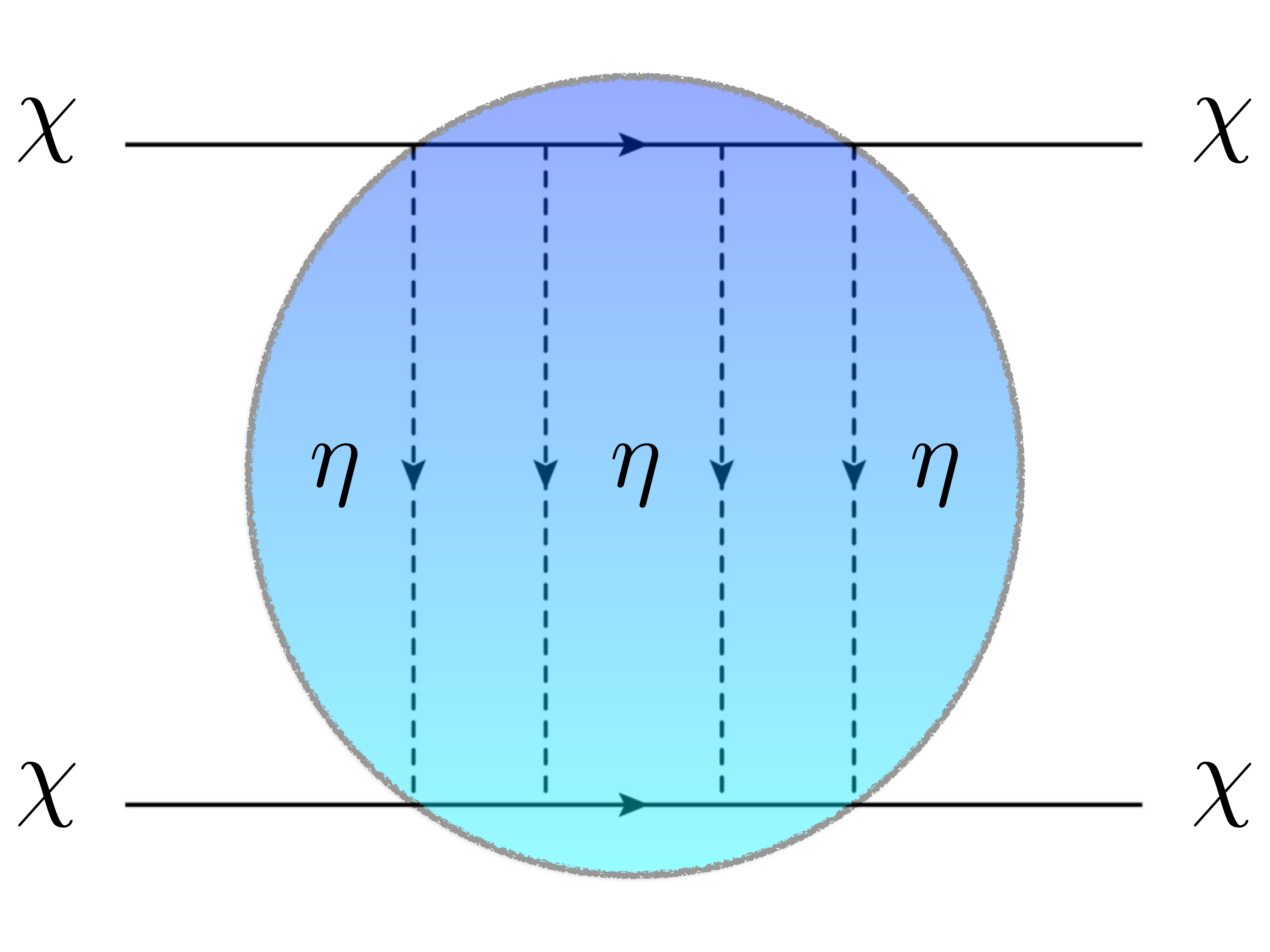} \label{fig:sketchSI}
\end{figure}

The dark sector contains a light particle, the Majoron $\eta$, coupled to Dark Matter. This opens the interesting possibility of a self-interacting Dark Matter candidate due to exchanges of the light particle via diagrams such as shown in Fig.~\ref{fig:sketchSI}.

Self-interacting Dark Matter could explain some of the issues encountered in simulations for small-scale structure formation which assume collisionless-DM~\cite{Oh:2010ea}, and typically predict too cuspy Dark Matter profiles. Self-interacting Dark Matter could explain the lack of satellites (although introducing baryons on the simulation seems to reduce inconsistencies~\cite{Blumenthal:1985qy,Gnedin:2004cx}) and more importantly the {\it too-big-to-fail} problem~\cite{Sawala:2010zw,BoylanKolchin:2011de} for $\sigma_{SI}/m_\chi \sim 0.1$-10 cm$^2$/g. 

Direct limits on self-interactions of Dark Matter are provided by lensing. From the renowned bullet cluster limit~\cite{Markevitch:2003at} to observations of other astrophysical objects, Dark Matter self interactions have been bounded in the range of $\sigma_{SI}/m_\chi < 1$ cm$^2$/g. Interestingly, there has been a recent claim of a measurement of self-interactions in the system Abell 3827~\cite{Massey:2015dkw} which lies above previous upper bounds. Note that this claim has been questioned by Ref.~\cite{Kahlhoefer:2015vua}, which propose modifications of the former analysis leading to limits similar to the bullet cluster's.

\begin{figure*}
\begin{center}
\includegraphics[width=0.98\textwidth]{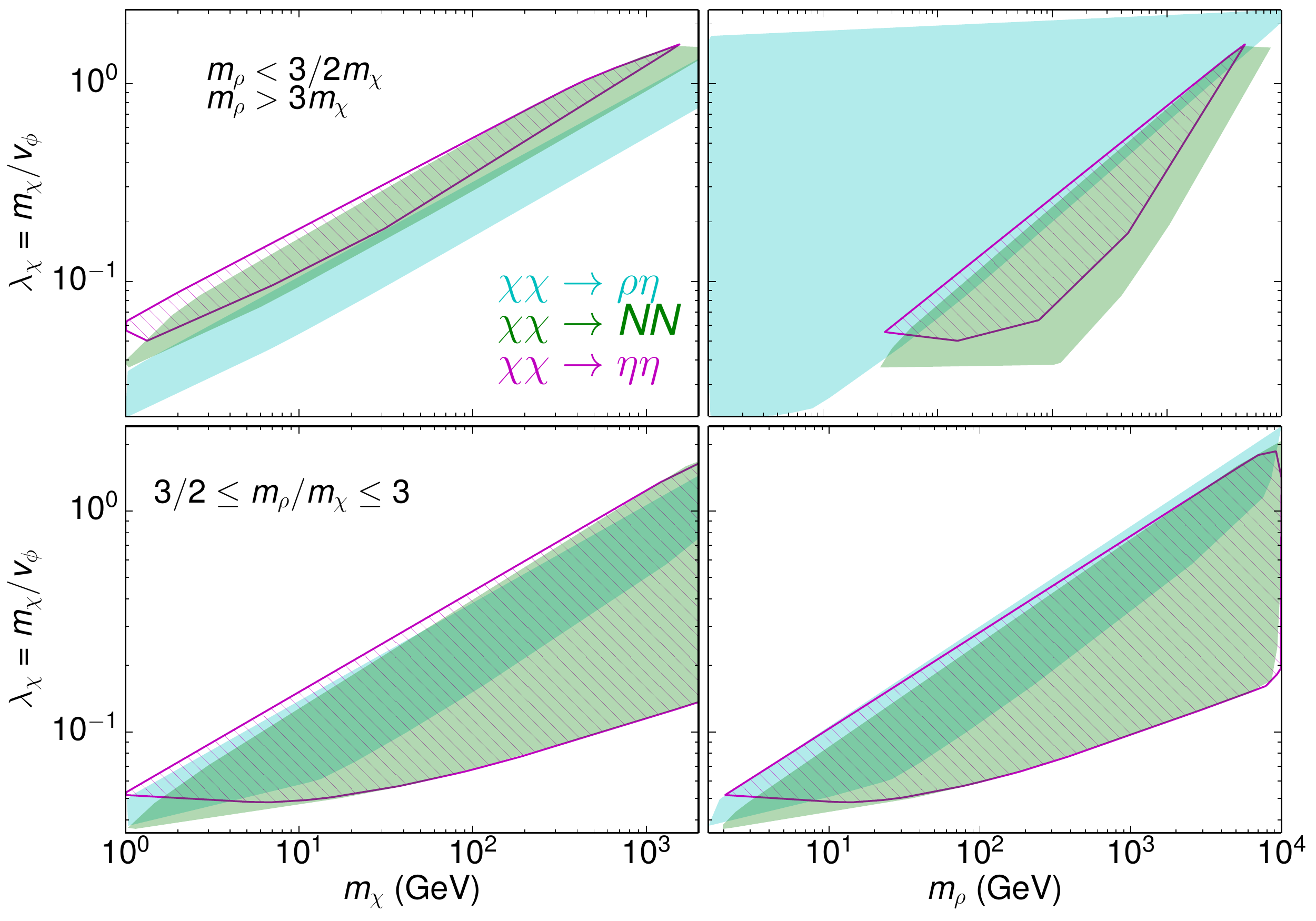} 
\end{center}
\label{fig:lambdachi}
\caption{Coupling between the dark scalars and Dark Matter $\lambda_\chi$ as a function of the Dark Matter mass $m_\chi$ (left) and heavy scalar mass $m_\rho$ (right). Outside the resonance (up) and on the resonance (down). The colors correspond to regions in parameter space in which the annihilation channel constitutes more than 60\% of the total cross section for $v=10^{-3}c$, as relevant for indirect detection.}
\end{figure*}

The effect of multiple exchanges of the light particle induces an Dark Matter effective potential between two dark particles $\chi$ of spin ${\bf s}$ at distance $r$
\bea
V_{eff} (r) = -\frac{\lambda_\chi^2 }{4 r^3 m_\chi^2} \, \left( 3 ({\bf s}_1. \hat{r})  ({\bf s}_2. \hat{r})- {\bf s}_1. {\bf s}_2\right) \ ,
\eea
where we neglected terms proportional to the (possible) Majoron mass. This potential is very singular at $r\to 0$ and requires regularization. The treatment for this case is quite involved, similar to a non-relativistic calculation of nucleon-nucleon interaction via the exchange of a light pion.
 
In the presence of these self-interactions, the annihilation cross section of a self-interacting Dark Matter would be modified respect to our discussion in Sec.~\ref{sec:relic}. Additional channels like $\chi\chi \to \chi \chi$ should be considered as they would be Sommerfeld-enhanced,
\bea
\sigma (\chi \chi \to \chi \chi) v_{rel} = S \, \frac{3 \lambda_\chi ^4}{64 \pi m_\chi^2} \ ,
\eea
where $S$ is the numerical factor due to Sommerfeld enhancement. In Ref.~\cite{Bellazzini:2013foa}, numerical calculations of $S$ at short distances were studied for this type of potential, finding that the enhancement could reach $S\sim 10^6$ for $v_{rel}=10^{-3}$. To estimate what values of $\lambda_\chi$ would lead to dominance of the self-interaction dynamics via a pseudo-scalar exchange, we follow Ref.~\cite{Archidiacono:2014nda} where the following bound is found
\bea
\lambda_\chi \gtrsim 0.6 \left( \frac{m_\chi}{\textrm{GeV}}\right)^{9/4} \ .
\label{conditionlambda}
\eea
Note that in our model the mass of the Dark Matter particle and $\lambda_\chi$ are related via the dark scalar vev $v_\phi$. We explored the range of these parameters leading to the correct relic abundance and the result is shown in Fig.~\ref{fig:lambdachi}. The estimate in~\ref{conditionlambda} corresponds to the upper-left corner in the left panel of this figure, away from the allowed region from the relic abundance constraint. 

Moreover, there are regions of the allowed parameter space where the scalar $\rho$ is light, and there will be exchanges of the $\rho$ similar to those in Fig.~\ref{fig:sketchSI} with $\rho$ as a mediator. As $\rho$ is a massive scalar mediator, the type of effective potential one would generate is Yukawa-type, with a less divergent behaviour than the pseudo-scalar.
 This case has been studied elsewhere, see e.g. Refs.~\cite{Iengo:2009ni,Cassel:2009wt,Cirelli:2007xd,LopezHonorez:2012kv,Lisanti:2016jxe}, and here we just quote the parametric dependence of the enhancement with the coupling,
 \bea
 S\sim \lambda_{\chi}^2/v_{rel} \ ,
 \eea 
 for attractive potentials and $m_\rho < m_{\chi}$. In the right panel of Fig.~\ref{fig:lambdachi} we show values of the coupling versus the mass of the mediator, finding in the region where the Sommerfeld enhancement could dominate, the self-interaction would compete with the s-wave annihilation into $\eta\rho$. 
 Note that a similar enhancement could happen in the channels of annihilation to right-handed neutrinos. Indeed, one could exchange light $\eta$ or $\rho$ mediators as in Fig.~\ref{fig:sketchSI}, but now between the Dark Matter particle and $N$. 

We conclude that the effects of self-interactions via the exchange of a pseudo-scalar mediator do not affect our model based on a naive estimate, but the effect of scalar exchanges and impact on the annihilation of Dark Matter into right-handed neutrinos deserve further study.

\section{Results}\label{sec:results}
In this section we show how the constraints discussed in the previous sections affect the parameter space of our model, described by $m_\chi, m_N, m_\rho$ and the Yukawa coupling 
$\lambda_{\chi}$, which fixes $v_\phi = m_\chi/\lambda_\chi$.

\begin{figure}[t!]

\begin{center}
\begin{tabular}{c}
\includegraphics[width=0.98\textwidth]{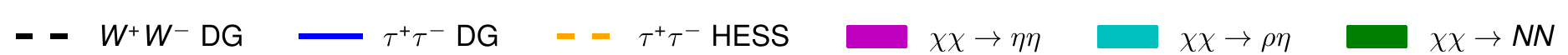} \\
\includegraphics[width=0.98\textwidth]{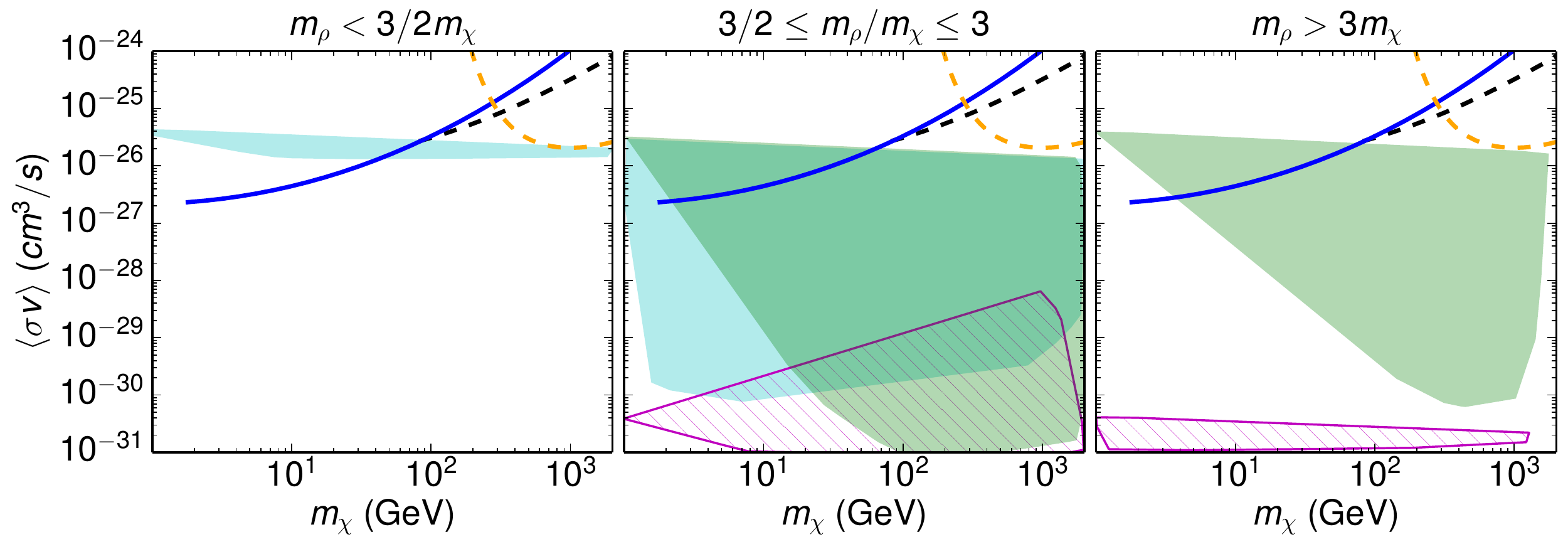} \\
\end{tabular}
\end{center}
\caption{Annihilation cross section at $v=10^{-3}c$ as a function of the Dark Matter mass as relevant for indirect detection. The color bands account for regions of parameter space in which the indicated annihilation channel provides more than 60\% of the total cross section. Also shown in this plot exclusion curves from FermiLAT dwarf galaxies (DG)~\cite{Ackermann:2015zua} and HESS galactic center (GC)~\cite{::2016jja}.}
\label{fig:s_ann}
\end{figure}

As discussed in Section~\ref{sec:ID} the annihilation product of the Dark Matter particle may lead to sizeable imprints on FermiLAT or HESS or the CMB due to the emission of photons and the re-ionization power of the products of the annihilation respectively. The annihilation channels that can lead a significant signature are to right-handed neutrinos $\chi \chi \to N N$ with $N N \to W^+W^-+$leptons, whenever $m_\chi > m_W$. 
A precise analysis of these decay channels would require a simulation of the photon spectrum from these cascade decays, such as performed in Ref.~\cite{Garcia-Cely:2016pse}.  Instead, we naively show the actual bounds for a 2 to 2 process in Fig.~\ref{fig:s_ann}.

Apart from signatures from gamma-rays, in our model neutrinos are typically produced in Dark Matter annihilation, leading to a flux from dense regions of Dark Matter or energy injection into the CMB.
Indeed, when the right-handed neutrino channel dominates, numerous neutrinos will be produced in the annihilations. IceCUBE can constrain the cross section to neutrinos measuring the flux from nearby Galaxies and Clusters (NG)~\cite{Aartsen:2013dxa}, the Galactic Halo (GH)~\cite{Aartsen:2014hva} and the Galactic Center (GC)~\cite{Aartsen:2015xej} and the CMB~\cite{Slatyer:2015jla} can constrain the annihilation cross section to neutrinos from the impact on re-ionization due to electroweak corrections. However, currently these probes lie three orders of magnitude above the model prediction, and thus cannot place a constrain on the model. 

It has been noted that for Dark Matter masses above 200 GeV annihilating into a pair of light neutrinos, 
Fermi-LAT data on gamma-rays sets the most stringent constraints on the annihilation cross-section 
\cite{Queiroz:2016zwd}.
In principle,  there would be similar limits in our scenario, but such  a heavy Dark Matter will also produce  $W$ bosons and leptons, which lead to stronger bounds discussed above.

In Fig.~\ref{fig:s_ann} we summarize these results in the plane of annihilation cross section into a particular final state versus the Dark Matter mass for the three kinematical regions of interest.
The colored contours correspond to regions with dominance of one channel in the relic abundance, either to dark scalars or right-handed neutrinos. As argued in Section~\ref{sec:ID} the annihilation products from the channels $\chi\chi\to \rho\eta$ and $\chi\chi\to \eta\eta$ cannot be constrained since the $\rho$ decays to two $\eta$'s, which are invisible.

Therefore the only limits that apply are those related to annihilation into right-handed neutrinos, which is suppressed by $m_N^2/m_\chi^2$. Promising signatures of these decays can be obtained when right-handed neutrinos undergo two-body decays in $W$ and charged leptons or neutrinos in association with a Higgs or a $Z$ boson, see Eq.~\ref{eq:WZh}, and hence restricted to $m_\chi > m_W $. On the other hand, for low Dark Matter mass only sterile neutrinos with $m_N <  m_W $ can be produced
and the dominant decay $N \to \nu \eta$ is unobservable, see Sec.~\ref{sec:ID}. Finally, note that the diagonal feature on the green region in the right panel is due to the fact that in the scan we have considered a minimum of 1 GeV for $m_N$.

From these results one can conclude that the model is currently unconstrained from indirect detection once other limits are taken into account, although the prospects for future experiments deserve a detailed study.

Notice that in the absence of the global $U(1)_{B-L}$ symmetry the conclusions could be quite different.
First,  the sterile neutrino mass would be independent of its coupling, $\lambda_N$,
 and there would not be any suppression of the annihilation cross section when  $m_N \ll m_\chi$.
 Moreover, since in our case the Majoron is a Goldstone boson,  it is expected to  be lighter than the  
 other scalars in the theory, and one can neglect its mass and assume that all possible annihilation 
 and decay channels into Majorons are always kinematically allowed. 
 However if the lepton number symmetry were explicitly broken, generically  the real and imaginary 
 components of $\phi$ would have similar masses, leading to cases where the channel into pseudo-scalars could be kinematically closed. 

 In summary, on the one hand the scenario we have considered with spontaneously broken 
 $U(1)_{B-L}$ is more  constrained than the one with explicit breaking, due to relation between sterile neutrino masses and couplings to $\phi$. On the other hand, if there was no symmetry, 
 we generically would  expect a heavier pseudo-scalar, which could lead to the closing of some invisible channels into Majorons.
In this case, constraints from the invisible Higgs decay width would be absent, Dark Matter would 
only  annihilate into sterile neutrinos, and the decay $N \rightarrow \nu \eta$  would not occur. Therefore
 limits from indirect searches, in particular the curve from Fermi-LAT dwarf galaxies on $\tau \tau$ shown in Fig.\ref{fig:s_ann}, would apply to low Dark Matter mass $\lesssim m_W$.

\section{Conclusions and outlook}\label{sec:concls}
In this paper we have studied a simple case connecting Dark Matter and the origin of neutrino masses, where the link to the Standard Model is dictated by a global $U(1)_{B-L}$ symmetry. In our model,
the dark sector   contains fermions, 
Dark Matter and right-handed neutrinos, 
and a complex scalar which plays the dual role of generating Majorana masses for the dark fermions and communicating with the Higgs via a Higgs portal coupling. 
 The stability of the Dark Matter fermion can be due to an additional dark sector symmetry, compositness or 
exotic lepton number.

After spontaneous electroweak and $U(1)_{B-L}$ symmetry breaking, the Higgs and dark scalar mix. This mixing is very constrained by bounds on the invisible width of the Higgs from the LHC and by LUX via the induced coupling of Dark Matter to the Higgs. 

We then focused on other aspects of the phenomenology of this model, assuming that the stable dark fermion constitutes the main component of Dark Matter in the Universe. Due to the presence of right-handed fermions and a complex scalar in the dark sector, there is an interplay between Dark Matter annihilation to both types of particles. Dark Matter annihilation to right-handed neutrinos could dominate at freeze-out provided the scalar is heavy. And, even when Dark Matter annihilation to Majorons dominated the dynamics at freeze-out, we found that the annihilation to heavy neutrinos could control today's indirect detection signatures.

Moreover, we found a very interesting phenomenology reaching from possible signatures at colliders via exotic Higgs decays, to effects on gamma-rays from right-handed neutrino production and decays to charged particles. In this paper, we did not try to accommodate a possible excess in the gamma-ray spectrum, instead used bounds from 2-to-2 scattering, adapted to our case in a relatively naive fashion. A proper study of the spectrum of gamma-rays in our model is beyond the scope of this paper, but certainly deserves further investigation since the estimated bounds are close to the WIMP thermal cross section.

Additionally, we noted that the presence of neutrinos in decay channels could be probed in the future via neutrino telescopes and more precise studies of the CMB, but that at the moment the limits are much weaker than any other annihilations involving charged particles. 

Finally, we briefly discussed the possibility of strong self-interactions of Dark Matter due to the exchange of the dark scalar. We found that Majoron exchange cannot dominate the Dark Matter dynamics, but the effect of exchanges of the dark scalar component deserves further study.

 \section*{Acknowledgements} 
We thank Pilar Hern\'andez, Laura Lopez-Honorez, Olga Mena, Sergio Palomares Ruiz, Roberto Ruiz de Austri, Jordi Salvad\'o and Sam Witte for illuminating discussions. 
ME specially thanks Antonia Abenza for inspiring and encouraging conversations.
This work has been partially supported by the European UnionÕs Horizon 2020 research and innovation programme under the Marie Sklodowska-Curie grant agreements No 674896 and 690575, 
by the Spanish MINECO under grants FPA2014-57816-P and  SEV-2014-0398,
and by Generalitat Valenciana grant PROMETEO/2014/050.
ME is supported by Spanish Grant FPU13/03111 of MECD. 
NR thanks the Department of Physics and Astronomy in the University of Sussex 
for the warm hospitality. The
work of VS is supported by the Science Technology and Facilities Council (STFC) under grant number ST/J000477/1.

 \bibliography{bibliography} 
 \end{document}